\documentclass[modern,times]{aastex61}

\usepackage{amsmath}          % AMS extended math package
\usepackage{natbib}           % package for citations
\usepackage{graphicx}         % EPS inclusion
\usepackage{xspace}           % gentle spacing after macros 
\usepackage{xifthen}          % improved it then construct for macro
\usepackage{mathrsfs}         % special math symbol for operators
\usepackage{txfonts}          % special math symbol

\bibpunct{(}{)}{;}{a}{}{,}    % match the AAs style

\providecommand*{\wn}{cm$^{-1}$\xspace}

\newcommand{\mrm}[1]{\ensuremath{\mathrm{#1}}}
\newcommand{\mcl}[3]{\multicolumn{#1}{#2}{#3}}

\newcommand{\pha}[1]{\phantom{#1}}

\newcommand{\oL}{\ensuremath{\hat{\mathscr{L}}}}
\newcommand{\oH}{\ensuremath{\tilde{H}}}
\newcommand{\oJ}{\ensuremath{\hat{J}}}

\newcommand{\fjk}[1]{\ensuremath{f_{#1}(J,k)}}
\newcommand{\txsum}[1]{\ensuremath{\textstyle{\sum\limits_{#1}}}}
\newcommand{\bra}[2][]{%
 \ifthenelse{\isempty{#1}}%
  {\ensuremath{\left\langle #2 \right\rvert}}%
  {\ensuremath{_{#1}\left\langle #2 \right\rvert}}}
\newcommand{\ket}[2][]{%
 \ifthenelse{\isempty{#1}}%
  {\ensuremath{\left\lvert #2 \right\rangle}}%
  {\ensuremath{\left\lvert #2 \right\rangle_{#1}}}}

\makeatletter
 \def\@dotsep{4.5}
\makeatother

\renewcommand{\thefootnote}{\fnsymbol{footnote}}
\newcommand{\hctn}{\mbox{HC\ensuremath{_3}N}\xspace}
\newcommand{\tnm}[1]{\tablenotemark{#1}}

\received{\today}
\revised{revision date}
\accepted{acceptance date}
\published{published date}
\submitjournal{\apjs}

\shorttitle{Accurate rest frequencies for \hctn}
\shortauthors{Bizzocchi et al.}

\begin{document}

\title{Rotational and high-resolution infrared spectrum of \hctn: global ro-vibrational 
      analysis and improved line catalogue for astrophysical observations}

\author{Luca Bizzocchi}
 \affiliation{Center for Astrochemical Studies, 
              Max-Planck-Institut f\"ur extraterrestrische Physik, \\
              Gie\ss enbachstr.~1, 85748 Garching bei M\"unchen (Germany)}

\author{Filippo Tamassia}
 \affiliation{Dipartimento di Chimica Industriale ``Toso Montanari'', 
              Universit\`a di Bologna, \\
              viale del Risorgimento~4, 40136 Bologna (Italy)}

\author{Jacob Laas}
\author{Barbara M. Giuliano}
 \affiliation{Center for Astrochemical Studies, 
              Max-Planck-Institut f\"ur extraterrestrische Physik, \\
              Gie\ss enbachstr.~1, 85748 Garching bei M\"unchen (Germany)}

\author{Claudio Degli Esposti}
\author{Luca Dore}
\author{Mattia Melosso}
 \affiliation{Dipartimento di Chimica ``G.~Ciamician'',
              Universit\`a di Bologna, \\
              via F.~Selmi 2, 40126 Bologna (Italy)}

\author{Elisabetta Can\`e}
 \affiliation{Dipartimento di Chimica Industriale ``Toso Montanari'', 
              Universit\`a di Bologna, \\
              viale del Risorgimento~4, 40136 Bologna (Italy)}

\author{Andrea Pietropolli Charmet}
 \affiliation{Dipartimento di Scienze Molecolari e Nanosistemi, 
              Universit\`a Ca'~Foscari Venezia, \\
              via Torino~155, 30172 Mestre (Italy)}

\author{Holger S.~P. M\"uller}
\author{Holger Spahn}
 \affiliation{I.Physikalisches Institut, 
              Universit\"at zu K\"oln, \\
              Z\"ulpicherstra\ss e 77, 50937 K\"oln (Germany)}

\author{Arnaud Belloche}
 \affiliation{Max-Planck-Institut f\"ur Radioastronomie, \\
              Auf dem H\"ugel~69, 53121 Bonn (Germany)}

\author{Paola Caselli}
 \affiliation{Center for Astrochemical Studies, 
              Max-Planck-Institut f\"ur extraterrestrische Physik, \\
              Gie\ss enbachstr.~1, 85748 Garching bei M\"unchen (Germany)}

\author{Karl M. Menten}
 \affiliation{Max-Planck-Institut f\"ur Radioastronomie, \\
              Auf dem H\"ugel~69, 53121 Bonn (Germany)}

\author{Robin T. Garrod}
  \affiliation{Departments of Chemistry and Astronomy, 
               University of Virginia, \\
              Charlottesville, VA 22904, USA}

\correspondingauthor{Luca Bizzocchi, Filippo Tamassia}
\email{bizzocchi@mpe.mpg.de;filippo.tamassia@unibo.it}

%_________________________________________________________________________________________

\begin{abstract}
\hctn is an ubiquitous molecule in interstellar environments, from external galaxies, 
to Galactic interstellar clouds, star forming regions, and planetary atmospheres. 
Observations of its rotational and vibrational transitions provide important information 
on the physical and chemical structure of the above environments.
We present the most complete global analysis of the spectroscopic data of \hctn. 
We have recorded the high-resolution infrared spectrum from~450 to~1350\,\wn, 
a region dominated by the intense $\nu_5$ and $\nu_6$ fundamental bands, located 
at~660 and~500\,\wn, respectively, and their associated hot bands. 
Pure rotational transitions in the ground and vibrationally excited states have been 
recorded in the millimetre and sub-millimetre regions in order to extend the frequency 
range so far considered in previous investigations.
All the transitions from the literature and from this work involving energy 
levels lower than 1000\,\wn have been fitted together to an effective Hamiltonian. 
Because of the presence of various anharmonic resonances, the Hamiltonian includes 
a number of interaction constants, in addition to the conventional rotational and 
vibrational $l$-type resonance terms.
The data set contains about~3400 ro-vibrational lines of 13~bands and some~1500 pure 
rotational lines belonging to~12 vibrational states. 
More than 120 spectroscopic constants have been determined directly from the fit, 
without any assumption deduced from theoretical calculations or comparisons with similar 
molecules. 
An extensive list of highly accurate rest frequencies has been produced to assist 
astronomical searches and data interpretation.
These improved data, have enabled a refined analysis of the ALMA observations towards
Sgr~B2(N2).
\end{abstract}

%\maketitle

%_________________________________________________________________________________________

\section{Introduction} \label{sec:intro}
\indent\indent
Cyanopolyynes, linear molecules with general formula HC$_{2n+1}$N, are among the 
most widespread species in astronomical environments.
The lightest members of this family are known to be primary constituents of the 
interstellar medium (ISM), as they have been identified in a variety of sources along 
the stellar evolutionary cycle.
Their chemistry is linked to that of the carbon chains, and has been successfully 
explained by chemical models using specific recipes for different physical conditions 
of the various phases of the star formation \citep[see][for a review]{Sakai-CR13-WCCC}.
In dark clouds, at early stages, when no protostar has yet ignited, cyanopolyynes are 
mainly generated by C$^+$ induced reactions, which proceed until most of the gas-phase 
carbon is locked in CO and then depleted onto dust grains
\citep[e.g.,][and references therein]{Loison-MNRAS14-CCs}.
Chains as long as HC$_{9}$N have been firmly identified in these environments, while the 
detection of HC$_{11}$N in TMC-1 \citep{Bell-ApJ97-HC11N} has been recently disputed 
by \citet{Loomis-MNRAS16-HC11N}.
At later stages, when the dust temperature rises ($T\sim 30-100$\,K) the chemistry is 
regenerated by carbon evaporation under hot core conditions and more chains are produced,
mainly by neutral--neutral reactions \citep{Sakai-ApJ08-CCs,Hassel-ApJ08-L1527}.

Cyanoacetylene (\hctn), the simplest cyanopolyyne, was first discovered towards the 
Galactic Centre by \citet{Turner-ApJ71-HC3N} and it has rapidly arisen as a major
astrophysical tracer.
Inside the Milky Way, \hctn is ubiquitous: it is abundant in starless cores 
\citep{Suzuki-ApJ92-DCCs}, massive star forming regions \citep{Li-ApJ12-MSFRs}, 
solar-type protostars \citep[e.g.][]{Jaber-AA17-IRAS16293}, carbon-rich circumstellar 
envelopes \citep{Decin-Nat10-IRC10216}, and post-AGB objects 
\citep{Pardo-ApJ04-CRL618,Wyro-ApJ03-CRL618}.
Moreover, its recent detection in proto-planetary discs 
\citep{Chapil-ApJ12-HC3N,Oberg-ApJ14-COMs} and comets \citep{Mumma-ARAA11-Comets} has
underlined its potential in an astrobiological context \citep{Oberg-Nat15-CN}, a role 
that had already been suggested many years ago \citep{Sanchez-Sci66-HC3N}.

\hctn has also been detected in external galaxies \citep[e.g.,][]{Mauer-AA90-HC3N}. 
\citet[e.g.,][]{Costa-AA10-HC3N} observed vibrationally excited \hctn towards NGC~4418 
and used it as a probe for the gas physical conditions in a source with intense infrared 
(IR) fields. 
The same galaxy was then re-investigated by \citet{Costa-AA15-NGC4418} who highlighted
the importance of \hctn in luminous infrared galaxies (LIRG).
In recent years, this species has been revealed in other extra-galactic sources: 
IC~342, M66 and NGC~660 \citep{Jiang-AA17-HC3N}, NGC~1097 \citep{Martin-AA15-NGC1097}, 
and in the nearby LIRG Mrk~231 by \citet{Aalto-AA12-Mrk231}.
\citet{Lind-AA11-HC3N} published a survey of 13 galaxies in which \hctn has been detected. 

Cyanoacetylene is also an important constituent of the atmosphere of the Saturn's major 
moon, Titan, where it was observed in the millimetre domain by \citet{Marten-Ica02-Titan}, 
in the infrared by the CIRS spectrometer on board the \textit{Cassini} spacecraft 
\citep{Couste-Ica07-Titan} and more recently, with ALMA \citep{Cord-ApJ14-Titan}.
Because of its importance, \hctn was included in a new 
astrobiological model of Titan's atmosphere \citep{Willacy-ApJ16-Titan}. 

The outstanding importance of \hctn for astrophysics and planetary sciences has
stimulated vast laboratory activity aimed at studying its spectroscopic properties.
The first observation of the pure rotational spectrum of cyanoacetylene dates back to 
the pioneering times of microwave spectroscopy \citep{Tyler-TFS63-MW} and the 
first precise measurements of its ground state transitions were performed in the 
centimetre (cm) region by \citet{deZafra-ApJ71-HC3N} using a molecular beam apparatus.
Later, the millimetre (mm) spectrum of \hctn was recorded by \citet{Cresw-JMS77-HC3N} 
and \citet{Mall-MP78-HC3N}, studies that were subsequently extended into the sub-millimetre 
(sub-mm) region by \citet{Chen-IJMW91-HC3N} and \citet{Yamada-ZN95-HC3N}.
A number of further laboratory studies were devoted to its vibrationally-excited
states spectra \citep{Mall-MP78-HC3N,Yamada-JMS86-HC3N,Mbosei-JMSt00-HC3N,Thorw-JMS00-HC3N},
and $l$-type transitions between the $l$-doublets of the bending excited states were
also recorded \citep{Laff-JMS68-HC3N,DeLeon-JCP85-HC3N}.

The investigation of the IR laboratory spectrum of \hctn started in the 70's and 
continued in the following decades with a number of low-resolution studies mainly aimed 
at the measurements of the absolute band intensities 
\citep{Uyemu-BCSJ74-HC3N,Uyemu-BCSJ82-HC3N,Khlifi-JMS90-HC3N,Khlifi-JMS92-HC3N}.
Rotationally resolved measurements were first performed in the 5\,$\mu$m spectral region 
by \citet{Yamada-ZN80-HC3N} and \citet{Yamada-ZN81-HC3N} with a IR diode laser
spectrometer.
The same authors also carried out a medium resolution study of the low energy portion of the
mid-IR spectrum \citep{Yamada-ZN86-HC3N}.

Later, \citet{Arie-JMS90-HC3N} published a detailed high-resolution investigation which 
covered the 450--730\,\wn range.
Several vibrational bands were identified, including the $\nu_5$ and $\nu_6$ fundamentals,
the $\nu_6 + \nu_7$ combination, plus a number of associated hot bands.
This study provided a listing of effective spectroscopic constants for the low-lying 
vibrational states of \hctn, that are the most interesting in the context of astrophysics.

As a matter of fact, observations of \hctn in the ISM very often involve excited 
vibrational states \cite[e.g.,][]{Peng-ApJ17-HC3N,Costa-AA15-NGC4418} and, in the context
of planetary sciences, a precise model of the observed infrared band profiles must 
include the associated hot bands \cite[e.g.,][]{Jolly-JMS07-HC3N}.
A detailed knowledge of the molecular ro-vibrational pattern is thus a prerequisite 
for a correct interpretation of the astronomical observations.
Because of the numerous perturbations that affect the rotation-vibration spectrum of \hctn,
a global analysis of the laboratory data including both pure-rotational and high-resolution 
IR measurements is necessary in order to derive a compact set of effective spectroscopic 
parameters without ambiguities, and to achieve spectral prediction of high accuracy.
To the best of our knowledge, unlike for the HCCC$^{15}$N species 
\citep{Fayt-JMSt04-HC3N15,Fayt-CP08-HC3N15}, there is no published global analysis for 
the main isotopologue.

Our purposes are: $i$) record new rotational and ro-vibrational spectra,
$ii$) perform a global fit of all the literature data with a careful treatment of the 
resonance effects, and $iii$) provide the best set of spectroscopic 
constants and a list of highly accurate rest-frequencies in the mm and IR spectral 
regions useful for astrophysical applications.
The structure of the paper is the following: in section~\ref{sec:exp} we describe the
experiments performed in various laboratories; in section~\ref{sec:theor} we give a summary
of the theory of the vibration-rotation spectra.
In the sections~\ref{sec:obs}--\ref{sec:anal} we describe our ensemble of spectroscopic data
and provide some details about the global analysis.
In section~\ref{sec:disc} we summarise the results, discuss the implications for astrophysics,
and report the rest frequency data list.
We present our conclusions in section~\ref{sec:conc}.

\section{Experiments} \label{sec:exp}
\indent\indent
A substantial amount of new spectroscopic data of \hctn have been collected in four 
laboratories located in Bologna, Cologne, and Munich.
The samples used for the measurements performed in Bologna and in Munich were prepared 
following the synthetic route described by \citet{Miller-SA67-NC6N}: propiolamide (Aldrich) 
was dehydrated with P$_4$O$_{10}$ at 225$^\circ$C under vacuum.
The gaseous products were collected in a trap kept at 77\,K and then purified by repeated
vacuum distillations to remove the volatile side-products (mainly NH$_3$).
The remaining white solid, composed by \hctn plus involatile polymers, was then directly
used for the spectroscopic measurements and could be stored at -25$^\circ$C over several
weeks without significant degradation.

The infrared spectra in the 450--1100\,\wn range were recorded in Bologna using a Bomem
DA3.002 Fourier-transform spectrometer equipped with a Globar source, a KBr
beam-splitter, and a liquid N$_2$-cooled HgCdTe detector.
Pathlengths of 0.16, 4 and 5\,m were employed.
Sample pressures ranging between~16 and~533\,Pa were used to record the spectra.
The resolution was generally~0.004\,\wn, except for the very weak $\nu_4$ band, which was
recorded at a lower resolution of~0.014\,\wn\@.
Several hundreds scans were co-added in order to improve the signal-to-noise (S/N) ratio
of the spectra.
The absolute calibration of the wavenumber axis was attained by referencing ro-vibrational
transitions of H$_2$O~\citep{Toth-JOSB91-H2O} and CO$_2$~\citep{Horne-JMS07-CO2}.
The accuracy of most line position measurements was estimated to be $5\times10^{-4}$\,\wn\@.

New mm-wave spectra in selected frequency intervals between 80 and 400\,GHz were 
observed in Bologna using a frequency-modulation (FM) mm-wave spectrometer whose details 
are reported elsewhere (see, e.g. \citealt{Bizz-ApJ16-ArH+}).
The Gunn oscillators, used as the primary radiation sources, were frequency-modulated at 
6\,kHz and second-harmonic ($2f$) detection was employed.
Further measurements of the submm-wave spectrum of \hctn in the 200--690\,GHz frequency 
range were carried out at the Centre for Astrochemical Studies (MPE Garching).
The complete description of this experimental apparatus is given in 
\citet{Bizz-AA17-HOCO+}: here the radiation source is a Virginia Diode multiplier chain 
driven by a centimetre-wave synthesizer.
FM at 15\,KHz and $2f$ detection was used.
In both laboratories, the spectra were recorded at room temperature, using static samples 
at a pressure of $\sim$ 0.5\,Pa.
The transition frequencies were recovered from a line-shape analysis of 
the spectral profile \citep{Dore-JMS03-proFFit} and their accuracy, estimated by repeated 
measurements, was in the 5--30\,kHz range, depending on the attained S/N\@.

The measurements performed in Cologne were carried out with left-over samples from 
previous studies \citep{Thorw-JMS00-HC3N,Yamada-ZN95-HC3N}.
Eight transitions pertaining to $\varv _5 = \varv _7 = 1$ ($J = 39, 41$) had remained 
unpublished by \citet{Thorw-JMS00-HC3N}.
Ground state transition frequencies were recorded in the 3\,mm region ($J = 8$ to $12$) 
to asses the best accuracy attainable for Doppler limited measurements of this molecule.
A 4\,m long single pass Pyrex glass cell equipped with PTFA windows was used for static 
measurements at room temperature and at pressures of 0.1\,Pa or lower. 
A backward-wave oscillator (BWO) based 3\,mm synthesizer AM-MSP~2 was employed as source, 
and a liquid He-cooled InSb bolometer as detector. 
Calibration measurements were made on the $J = 1 - 0$, CO line whose frequency is known 
to an accuracy of 0.5\,kHz from sub-Doppler measurements \citep{Winn-JMS97-CO}.
After adjustment, this line was measured in Doppler regime with a precision of $\sim$ 
2\,kHz.

Further measurements were made using the Cologne Terahertz Spectrometer (CTS, see 
\citealt{Winn-VB05-CTS} for a detailed description of the apparatus).
Room-temperature static samples at pressures of 0.1\,Pa were employed for stronger lines, 
and up to about 1\,Pa for the weaker ones.
A few lines were recorded around 610\,GHz, 800\,GHz and 900\,GHz.
These measurements were aimed at improving the data set for some vibrationally excited 
states not investigated by \citet{Thorw-JMS00-HC3N}, and at achieving a general spectral 
coverage extending beyond $J = 100$.
Measurement accuracies for isolated lines with very symmetric shape ranged from 5\,kHz to 
mostly $10-20$\,kHz. 
Weaker or less symmetric lines or lines close to others were given larger uncertainties. 
The measurements and the accuracies are similar to DC$_3$N described in 
\citet{Spahn-CP08-DC3N}.

\section{Theory} \label{sec:theor}

%%%%% TABLE ORDER OF MAGNITUDE H TERMS
\begin{deluxetable}{c @{\hskip 2em} cc}[b!]
 \tablecaption{Order-of-magnitude classification of the resonance operators 
               \label{tab:odg}}
 \tablehead{
  \colhead{terms} & \mcl{2}{c}{order of magnitude} \\
  \cline{2-3} \\[-2.5ex]
  \colhead{} & \colhead{$\kappa^{m+n-2}\omega_\text{vib}$} & 
               \colhead{$\kappa^{2m+2n-5}\omega_\text{vib}$} \\[-0.5ex]
  \colhead{} & \colhead{``exact'' resonance}  & \colhead{close interacting levels}
 }
 \startdata
  $\oH_{30}$           &  $\kappa\omega_\text{vib}$   &  $\kappa\omega_\text{vib}$      \\
  $\oH_{31},\oH_{40}$  &  $\kappa^2\omega_\text{vib}$ &  $\kappa^3\omega_\text{vib}$    \\
  $\oH_{32},\oH_{50}$  &  $\kappa^3\omega_\text{vib}$ &  $\kappa^5\omega_\text{vib}$    \\
  $\oH_{33},\oH_{42}$  &  $\kappa^4\omega_\text{vib}$ &  $\kappa^7\omega_\text{vib}$    \\
  $\oH_{52},\oH_{34}$  &  $\kappa^5\omega_\text{vib}$ &  $\kappa^9\omega_\text{vib}$    \\
  $\oH_{44}$           &  $\kappa^6\omega_\text{vib}$ &  $\kappa^{11}\omega_\text{vib}$ \\
  $\oH_{54}$           &  $\kappa^7\omega_\text{vib}$ &  $\kappa^{13}\omega_\text{vib}$ \\[0.5ex]
 \enddata 
\end{deluxetable}

\subsection{Notation for states and wave-functions} \label{sec:wavef}
\indent\indent
The analysis presented in this paper involves, as far as the ro-vibrational data are 
concerned, transitions arising from the ground state and four vibrationally-excited states 
located below 1100\,\wn: $\varv_7$ (C$-$CN bend ), $\varv_6$ (CCC bend), 
$\varv_5$ (H$-$CC bend), and $\varv_4$ (C$-$C stretch).
All the other vibrational modes are not considered, thus a given vibrational state can 
be labelled using the notation $(\varv_4,\varv_5^{l_5},\varv_6^{l_6},\varv_7^{l_7})_{e/f}$, 
where $l_t$ quantum numbers label the vibrational angular momentum associated to each 
$\varv_t$ bending mode.
The $e/f$ subscripts indicate the parity of the symmetrised wave-functions following the 
usual convention for linear molecules \citep{Brown-JMS75-ef}.
When there is no ambiguity, the simplified notation $(l_5,l_6,l_7)_{e/f}$ will be 
used in the text to identify the different sub-levels of a bending state.  

The ro-vibrational wave-functions are represented by the ket
$\ket[e/f]{\varv_4,\varv_5^{l_5},\varv_6^{l_6},\varv_7^{l_7};J,k}$.
The vibrational part is expressed as a product of one- and two-dimensional harmonic 
oscillator wave-functions, while the rotational part is the symmetric-top wave-function
where the angular quantum number $k$ is subjected to the constraint 
$k = l_5 + l_6 + l_7$\@.
Symmetry-adapted basis functions are obtained by the following Wang-type linear 
combinations \citep{Yamada-JMS85-YBA}
\begin{subequations} \label{eq:Wang}
\begin{align}
 \left\lvert \varv_4,\varv_5^{l_5}\right.,&\left.\varv_6^{l_6},\varv_7^{l_7};J,k\right\rangle_{e/f}  \notag \\
   & = \frac{1}{\sqrt{2}} \left\{
     \ket{\varv_4,\varv_5^{l_5},\varv_6^{l_6},\varv_7^{l_7};J,k} \pm (-1)^k
     \ket{\varv_4,\varv_5^{-l_5},\varv_6^{-l_6},\varv_7^{-l_7};J,-k} 
                          \right\} \,,   \\
 \left\lvert \varv_4,0^0\right.,&\left.0^0,0^0;J,0\right\rangle_{e} = \ket{\varv_4,0^0,0^0,0^0;J,0} \,.
\end{align}
\end{subequations}
The upper and lower signs ($\pm$) correspond to $e$ and $f$ wave-functions, respectively.
For $\Sigma$ states ($k = 0$), the first non-zero $l_t$ is chosen positive.
Note that the omission of the $e/f$ label indicates unsymmetrised wave-functions.

\subsection{Ro-vibrational Hamiltonian} \label{sec:Hvr}
\indent\indent
The observed transition frequencies are expressed as differences between ro-vibrational
energy eigenvalues; these are computed using an effective Hamiltonian adapted for a 
linear molecule:
\begin{equation} \label{eq:H}
 \oH = \oH_\text{vr} + \oH_{l\text{-type}} + \oH_\text{res} \,,
\end{equation}
where $\oH_\text{vr}$ represents the ro-vibrational energy including centrifugal
distortion, $\oH_{l\text{-type}}$ is the $l$-type interaction energy among the $l$
sub-levels of excited bending states, and $\oH_\text{res}$ is the contribution due
to the ro-vibrational resonances between accidentally quasi-degenerate states.

The Hamiltonian matrix is set up using unsymmetrised ro-vibrational basis functions;
it is then factorised, and symmetrised using Eqs.~\eqref{eq:Wang}.
The matrix elements of the effective Hamiltonian are expressed using the formalism 
first introduced by \citet{Yamada-JMS85-YBA} and already employed for analysis of the 
ro-vibrational spectra of several carbon chains with multiple bending vibrations
(see e.g., \citealt{Bizz-JMS05-HC5N}).
Here, the following shorthand will be used to simplify the notation:
\begin{subequations}
\begin{align} \label{eq:fjk}
  &\fjk{0} = J(J + 1) - k^2 \,,  \\
  &\fjk{\pm n} = \prod\limits_{p=1}^n\: J(J + 1) - [k \pm (p - 1)](k \pm p) \,.
\end{align}
\end{subequations}

The $\oH_\text{vr}$ term of the Hamiltonian is purely diagonal in all the $\varv$ and $k$
quantum numbers.
It has the form
\begin{align} \label{eq:diag}
 \bra{l_5,l_6,l_7;k}& \oH_\text{vr} \ket{l_5,l_6,l_7;k} =
      \txsum{t} x_{L(tt)}l_t^{\,2} + \txsum{t\neq 7} x_{L(t7)}l_tl_7
                                + \txsum{t} y_{L(tt)}l_t^{\,4}              \notag \\
   &+ \left\{ B_v + \txsum{t} d_{JL(tt)}l_t^{\,2}
                  + \txsum{t\neq 7} d_{JL(t7)}l_tl_7 \right\} \fjk{0}    \notag \\
   &- \left\{ D_v + \txsum{t} h_{JL(tt)}l_t^{\,2}
                  + \txsum{t\neq 7} h_{JL(t7)}l_tl_7 \right\} \fjk{0}^2  \notag \\
   &+ \left\{ H_v + \txsum{t} l_{JL(tt)}l_t^{\,2} \right\} \fjk{0}^3 \,.
\end{align}

The $\oH_{l\text{-type}}$ term of the Hamiltonian is also diagonal in $\varv$, but it 
features contributions which are off-diagonal in the quantum numbers $l_t$ and with
$\Delta k = 0, \pm 2, \pm 4$\@.

The vibrational $l$-type doubling terms with $\Delta k = 0$ have the general formula
\begin{multline} \label{eq:dk0}
 \bra{l_t\pm 2,l_{t'}\mp 2;k} \oH_{l\text{-type}} \ket{l_t,l_{t'};k} =
     \tfrac{1}{4}\left\{ r_{tt'} + r_{tt'J}J(J+1) + r_{tt'JJ}J^2(J+1)^2 \right\}  \\
     \times\sqrt{(\varv_t\mp l_t)(\varv_t\pm l_t + 2)(\varv_{t'}\mp l_{t'} + 2)
                 (\varv_{t'}\pm l_{t'})} \,.
\end{multline}

The rotational $l$-type resonance terms with $\Delta k = \pm 2$ are expressed by
\begin{multline} \label{eq:dk2a}
 \bra{l_t\pm 2;k\pm 2} \oH_{l\text{-type}} \ket{l_t;k} =
     \tfrac{1}{4}\left\{ q_t + q_{tJ}J(J+1) + q_{tJJ}J^2(J+1)^2 \right\}  \\
     \times\sqrt{(\varv_t\mp l_t)(\varv_t\pm l_t + 2)} \sqrt{\fjk{\pm 2}} \,,
\end{multline}
\begin{multline} \label{eq:dk2b}
 \bra{l_t\mp 2,l_{t'}\pm 4;k\pm 2} \oH_{l\text{-type}} \ket{l_t,l_{t'};k} =
     \tfrac{1}{8} q_{tt't'} \{ (\varv_t\mp l_t + 2)(\varv_t\pm l_t)(\varv_{t'}\mp l_{t'})  \\
     (\varv_{t'}\pm l_{t'} + 2)(\varv_{t'}\mp l_{t'} - 2)(\varv_{t'}\pm l_{t'} + 4) \}^{1/2}
     \sqrt{\fjk{\pm 2}}  \,.
\end{multline}

The terms relative to $\Delta k = \pm 4$ are
\begin{multline} \label{eq:dk4a}
 \bra{l_t\pm 4;k\pm 4} \oH_{l\text{-type}} \ket{l_t;k} =
     \tfrac{1}{4} u_{tt} \{ (\varv_t\mp l_t)(\varv_t\pm l_t + 2)
     (\varv_t\mp l_t - 2)(\varv_t\pm l_t + 4) \}^{1/2} \sqrt{\fjk{\pm 4}} \,,
\end{multline}
\begin{multline} \label{eq:dk4b}
 \bra{l_t\pm 2,l_{t'}\pm 2;k\pm 4} \oH_{l\text{-type}} \ket{l_t,l_{t'};k} =
     \tfrac{1}{4} u_{tt'} \{ (\varv_{t'}\mp l_{t'})(\varv_{t'}\pm l_{t'} + 2) \\
     (\varv_t\mp l_t)(\varv_t\pm l_t + 2) \}^{1/2} \sqrt{\fjk{\pm 4}} \,.
\end{multline}

Following \citet{Wagner-JMS93-HC15NO}, the terms of the effective Hamiltonian for the
ro-vibrational resonances can be written as
\begin{equation} \label{eq:genres}
  \oH_\text{res} = \sum_{m,n} C_{mn} \oL^m \oJ^n \,,
\end{equation}
where $C_{mn}$ is the resonance coefficient, $m$ the total degree in the vibrational
ladder operators $\oL^{\pm}$, and $n$ the total degree in the rotational angular momentum
operators $\oJ$ \citep{Wagner-JMS93-HC15NO,Okab-JMS99-FC3N}. 
Here we want to discuss briefly the order of magnitude of the terms connecting the 
interacting states. 
Given the complexity of the \hctn resonance network, order-of-magnitude considerations
are very useful to assess which terms matter in a given range of energy and quantum 
numbers and which can be safely neglected.

The order of magnitude of the terms in the rotation-vibration Hamiltonian is usually 
expressed as the Born-Oppenheimer expansion parameter, 
$\kappa = (m_e/m_n)^{1/4}$, where $m_e$ and $m_n$ are the electronic and nuclear masses, 
respectively \citep{Oka-JCP67-A1A2}.
For ro-vibrational spectroscopy applications, a suitable estimate is 
$\kappa = (B/\omega_\text{vib})^{1/2}$, where $\omega_\text{vib}$ is a typical harmonic 
vibrational frequency, and $B$ is the rotational constant \citep{Nielsen-RMP51-VibRot}.
For \hctn, $\kappa\simeq 1/56$, taking $B\sim 0.15$\,\wn and 
$\omega_\text{vib}\sim 500$\,\wn\@.
Following \citet{A&W-MR85-VibRot}, the general expression for the order of magnitude of 
the effective Hamiltonian element, $\oH_{mn}$, can be written as
\begin{equation} \label{eq:ex-res}
  \oH_{mn} \approx r^m J^n \kappa^{m+2n-2} \omega_\text{vib} \,,
\end{equation}
where $r$ is either the vibrational coordinate $q$ or the vibrational momentum $p$, and
$J$ is the rotational quantum number.
This latter dependence accounts for resonances that involve rotational operators: 
these terms can be neglected at low or moderate values of $J$ but may become important 
as $J$ increases.

For low vibrational quantum numbers, $r\simeq 1$ and if one assumes 
$J\simeq\kappa^{-1}\simeq 56$, the order of magnitude of the $\oH_{mn}$ contribution is 
$\kappa^{m+n-2}\omega_\text{vib}$ for exact resonances.
For less close resonances, the contribution to the ro-vibrational energy of the matrix
element $H_{mn}$ can be estimated through the $2^\text{nd}$ order perturbation formula
\begin{equation} \label{eq:close-res}
  E \simeq \frac{H_{mn}^2}{\Delta} \simeq \kappa^{2m+2n-4} 
    \left(\frac{\omega_\text{vib}}{\Delta}\right) \omega_\text{vib} \,.
\end{equation}
The quantity $\Delta$ in the denominator of Eq.~\eqref{eq:close-res} is the energy
difference between the two vibrational levels.
For interacting states, whose energy difference is $\Delta\simeq 10$\,\wn, one can 
assume $\omega_\text{vib}/\Delta \simeq \kappa^{-1}$. 
The contribution of $\oH_{mn}$ to the ro-vibrational energy is then 
$\kappa^{2m+2n-5}\omega_\text{vib}$.

The order-of-magnitude classification of the various Hamiltonian terms is summarised in
Table~\ref{tab:odg} where, in the first column, we listed all the resonance operators 
which may be relevant for the analysis of the spectra described in this paper.
It is important to notice that the overall rank of the individual terms is preserved in 
the two cases described. 
Only the power of $\kappa$ diverges more rapidly in close-resonance cases. 
This implies that, once we choose the cutting threshold for the power of $\kappa$ for the 
terms to be considered in the analysis, more interactions have to be taken into account 
in the case of ``exact'' resonances with respect to the close resonance situation.

The resonance network present in the energy level manifold of \hctn below 1000\,\wn had
already been described and partially analysed by \citet{Yamada-JMS86-HC3N}. 
However, given the higher level of details of our investigation, a number of extra terms 
have been evaluated.
As a general guideline, energy contributions of order higher than 
$\kappa^5\omega_\text{vib}$ ($<0.03$\,MHz) can be safely neglected. 

Our analysis indicates that $\oH_{30}$, $\oH_{31}$, and $\oH_{40}$ must be considered, 
$\oH_{32}$ and $\oH_{50}$ can be important for close interacting levels.
On the other hand, $\oH_{42}$ and $\oH_{52}$ might produce only minor effects on very 
close resonances and their importance can be significantly enhanced for high-$\varv$, 
high-$J$ levels.
In the following Eqs.~\eqref{eq:res30-466}--\eqref{eq:res42b}, we list all the resonance 
terms included in the spectral analysis.

The cubic anharmonic interactions of the ($\varv_4$,$\varv_5$,$\varv_6$,$\varv_7$) state 
with ($\varv_4+1$, $\varv_5$, $\varv_6-2$, $\varv_7$), and 
($\varv_4+1$,$\varv_5-1$,$\varv_6$,$\varv_7-1$) states are expressed by
\begin{multline} \label{eq:res30-466}
  \bra{\varv_4,\varv_5^{l_5},\varv_6^{l_6},\varv_7^{l_7};J,k}
    \oH_{30} + \oH_{32}
  \ket{\varv_4+1,\varv_5^{l_5},(\varv_6-2)^{l_6},\varv_7^{l_7};J,k} \\
    = \sqrt{2}\left[(\varv_4 + 1)\,(\varv_6 + l_6)(\varv_6 - l_6)\right]^{1/2}
      \left\{C_{30}^{(466)} + C_{32}^{(466J)} J(J + 1)\right\} \,,
\end{multline}
\begin{multline} \label{eq:res30-457}
  \bra{\varv_4,\varv_5^{l_5},\varv_6^{l_6},\varv_7^{l_7};J,k}
    \oH_{30} + \oH_{32}
  \ket{\varv_4+1,(\varv_5-1)^{l_5\pm 1},\varv_6^{l_6},(\varv_7-1)^{l_7\mp 1};J,k} \\
    = \frac{\sqrt{2}}{2}\left[(\varv_4 + 1)\,(\varv_5 \mp l_5)(\varv_7 \pm l_7)\right]^{1/2}
      \left\{C_{30}^{(457)} + C_{32}^{(457J)} J(J + 1)\right\} \,.
\end{multline}

The quartic anharmonic interaction coupling the ($\varv_5$,$\varv_7$) and 
($\varv_5+1$, $\varv_7-3$) states is given by
\begin{align} \label{eq:res40-5777}
  \bra{\varv_5^{l_5},\varv_7^{l_7};J,k} &\oH_{40} + \oH_{42} 
  \ket{(\varv_5+1)^{l_5\pm 1},(\varv_7-3)^{l_7\mp 1};J,k} \notag \\
    = &\frac{1}{4}\left[(\varv_5 \pm l_5 + 2)(\varv_7 \pm l_7)(\varv_7 \mp l_7)
                        (\varv_7 \pm l_7 - 2) \right]^{1/2}   \notag \\
      &\left\{C_{40}^{(5777)} + C_{42}^{(5777J)} J(J + 1)\right\} \,.
\end{align}

The quintic anharmonic resonance between the ($\varv_4$,$\varv_7$) and 
($\varv_4+1$,$\varv_7-4$) states is
\begin{align} \label{eq:res50-47777}
  \bra{\varv_4,\varv_7^{l_7};J,k} &\oH_{50} + \oH_{52} 
  \ket{(\varv_4+1),(\varv_7-4)^{l_7};J,k} \notag \\
    = &\frac{\sqrt{2}}{2}\left[(\varv_4 + 1)\,(\varv_7 + l_7 - 2)(\varv_7 - l_7 - 2)
                               (\varv_7 + l_7)(\varv_7 - l_7) \right]^{1/2} \notag \\
      &\left\{C_{50}^{(47777)} + C_{52}^{(47777J)} J(J + 1)\right\} \,.
\end{align}

In association with the classic quartic anharmonic resonance, we have taken into account 
two further terms generated by the $\oH_{42}$ Hamiltonian and that are off-diagonal both
in $\varv$ and in $l$. 
They are
\begin{align} \label{eq:res42a}
  \bra{\varv_5^{l_5},\varv_7^{l_7};J,k} \oH_{42} &
  \ket{(\varv_5+1)^{l_5\pm 1},(\varv_7-3)^{l_7\pm 1};J,k \pm 2} \notag \\
    = &\frac{1}{4}\left[(\varv_5 \pm l_5 + 2)(\varv_7 + l_7)(\varv_7 - l_7)
                        (\varv_7 \mp l_7 + 2) \right]^{1/2}  \notag \\
      &\times C_{42a}^{(5777)} \sqrt{\fjk{\pm 2}} \,,
\end{align}
\begin{align} \label{eq:res42b}
  \bra{\varv_5^{l_5},\varv_7^{l_7};J,k} \oH_{42} &
  \ket{(\varv_5+1)^{l_5\pm 1},(\varv_7-3)^{l_7\mp 3};J,k \mp 2} \notag \\
    = &\frac{1}{4}\left[(\varv_5 \pm l_5 + 2)(\varv_7 \pm l_7)(\varv_7 \pm l_7 - 2)
                        (\varv_7 \mp l_7 - 4) \right]^{1/2} \notag \\
      &\times C_{42b}^{(5777)} \sqrt{\fjk{\mp 2}} \,.
\end{align}

Other possible couplings are: the quartic anharmonic interaction between the 
($\varv_5$,$\varv_6$,$\varv_7$) and ($\varv_5+1$,$\varv_6-2$,$\varv_7+1$) states generated 
by $\oH_{40}$, plus the $\oH_{31}$ Coriolis-type resonances that couple the states 
($\varv_6$,$\varv_7$) and ($\varv_6+1$,$\varv_7-2$), and ($\varv_5$,$\varv_6$,$\varv_7$) 
and ($\varv_5+1$,$\varv_6-1$,$\varv_7-1$).
These interactions were found to be important in the global fit of the infrared and 
rotational spectra of the HCCC$^{15}$N isotopologue
\citep{Fayt-JMSt04-HC3N15,Fayt-CP08-HC3N15}.
We have tested these terms in the \hctn analysis, but they produce only minor effects 
and the corresponding coefficients were poorly determined.
Hence, these interactions were not considered in the final fit 
(see also \S~\ref{sec:anal:iso} and \S~\ref{sec:disc:gen}).

%%%% FIGURE ENERGY LEVELS
\begin{figure}[b!]
 \centering
 \includegraphics[angle=0,width=0.6\textwidth]{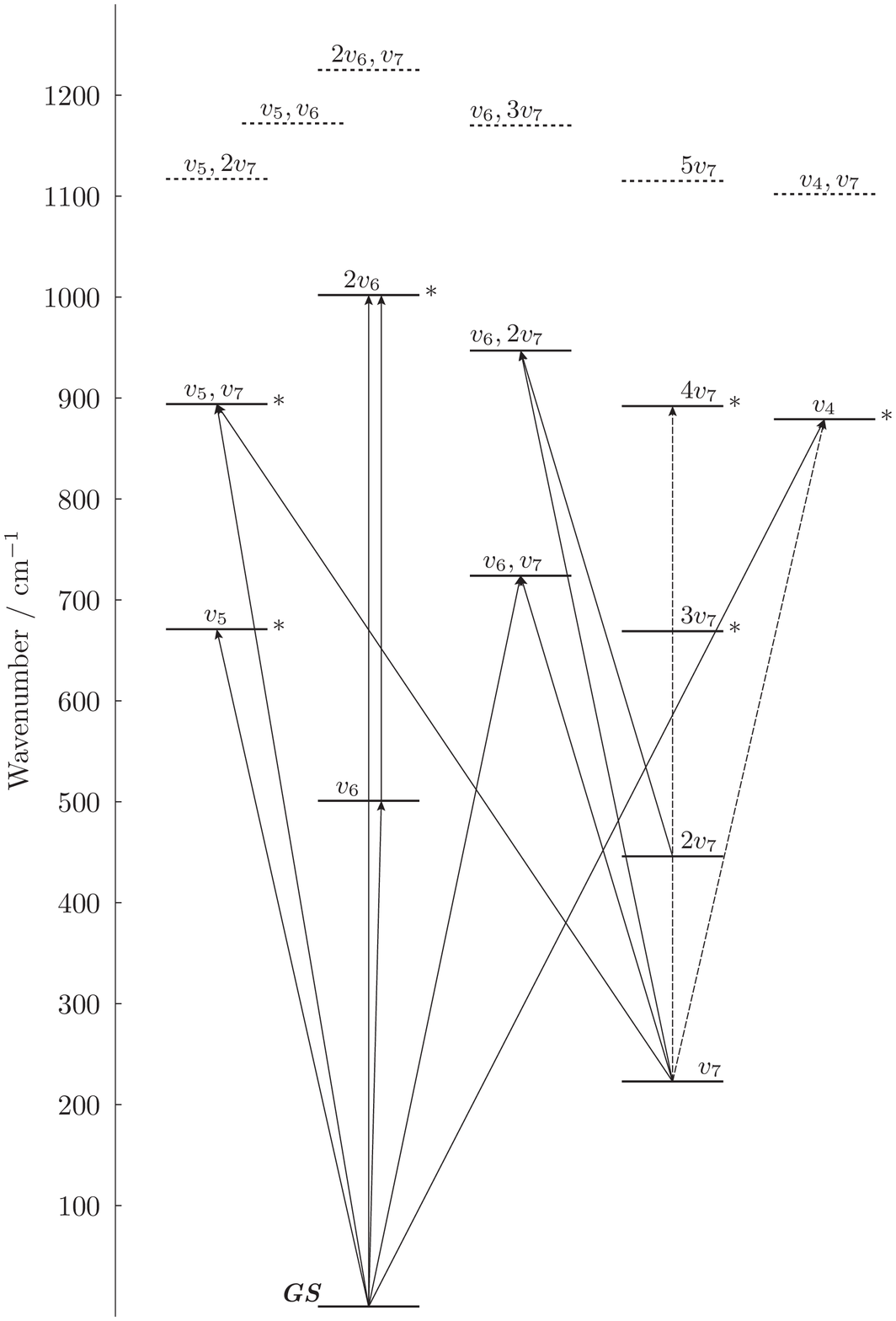}
 \caption{Vibrational energy level diagram of \hctn up to $\sim$ 1300\,\wn.
          The states are labelled in a compact manner with the indication of the excited
          quanta.
          The investigated levels are plotted with solid horizontal lines, and the arrows
          show the 13~IR bands analysed.
          Interacting states are marked with an asterisk.
          The two dashed arrows indicate the perturbation enhanced bands.}
  \label{fig:lev-scheme}
\end{figure}

\section{Observed spectra} \label{sec:obs}

%%%% FIGURE IR SPECTRUM
\begin{figure*}[t!]
  \centering
  \includegraphics[angle=0,width=0.90\textwidth]{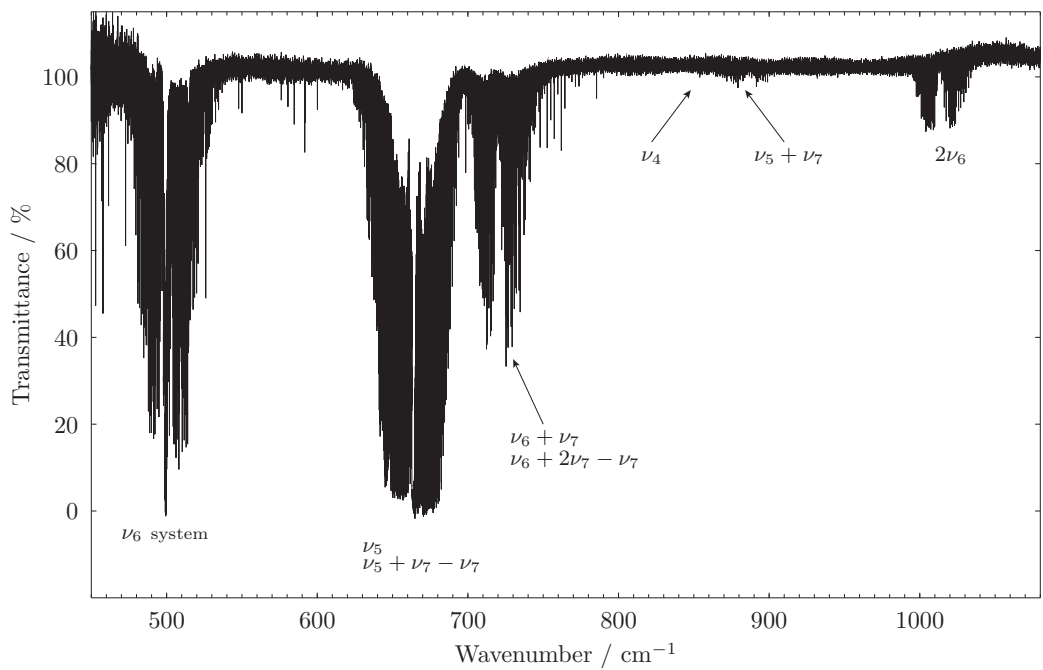}
  \caption{Overview of the infrared spectrum of HC$_3$N in the 450--1100\,\wn\ region. 
           The analysed bands are indicated. 
           The $\nu_6$ system includes the $\nu_6+\nu_7-\nu_7$, $2\nu_6-\nu_6$, and 
           $\nu_6+2\nu_7-2\nu_7$ hot bands.
           Recording conditions: $T=298$\,K, $P=67$\,Pa, $L_\text{path}=4$\,m, 880~scans, 
           unapodised resolution 0.004\,\wn\@.
          }
  \label{fig:overview}
\end{figure*}

%%%%% TABLE IR DATA
\begin{deluxetable*}{l cccccc}[h!]
 \tablecaption{Ro-vibrational bands of \hctn analysed in this work
               \label{tab:summIR}}
 \tabletypesize{\footnotesize}
 \tablehead{
  \colhead{Band}         &
  \colhead{Sub-bands}    &
  \colhead{Wav. range}   &
  \colhead{$P,Q,R$}      &
  \colhead{No. of lines} &
  \colhead{$\sigma_i$\tnm{a}}   \\
  \colhead{}             &
  \colhead{}             &
  \colhead{[\wn]}        &
  \colhead{($J_\mrm{min}-J_\mrm{max}$)} &
  \colhead{}             &
  \colhead{[$10^{-3}$\,\wn]}
 }
 \startdata
  $\nu_6$               & $\Pi-\Sigma^+$            &  477 --  523  &  $P(3-77)$, $Q(3-83)$, $R(2-72)$    &  218 &  0.5 \\[-0.5ex]
  $\nu_6+\nu_7-\nu_7$   & $(\Sigma^\pm,\Delta)-\Pi$ &  475 --  528  &  $P(2-89)$, $Q(3-89)$, $R(2-91)$    &  647 &  0.5 \\[-0.5ex]
  $2\nu_6-\nu_6$        & $(\Sigma^+,\Delta)-\Pi$   &  479 --  536  &  $P(1-73)$, $Q(18-73)$, $R(1-79)$   &  435 &  0.5 \\[-0.5ex]
  $\nu_6+2\nu_7-2\nu_7$ & $\Pi-(\Sigma^+,\Delta)$   &  474 --  524  &  $P(3-82)$, $R(3-78)$               &  339 &  0.5 \\[-0.5ex]
  $\nu_5$               & $\Pi-\Sigma^+$            &  632 --  696  &  $P(2-103)$, $Q(18-78)$, $R(0-105)$ &  264 &  0.5 \\[-0.5ex]
  $\nu_5+\nu_7-\nu_7$   & $(\Sigma^\pm,\Delta)-\Pi$ &  641 --  691  &  $P(1-70)$, $Q(4-59)$, $R(1-91)$    &  623 &  1.0 \\[-0.5ex]
  $\nu_4-\nu_7$ $\ast$  & $\Sigma^+-\Pi$            &  617 --  650  &  $P(9-65)$, $Q(8-69)$, $R(3-34)$    &  118 &  1.0 \\[-0.5ex]
  $4\nu_7-\nu_7$ $\ast$ & $(\Sigma^+,\Delta)-\Pi$   &  648 --  672  &  $P(24-50)$, $R(25-48)$             &   19 &  1.0 \\[-0.5ex]
  $\nu_6+\nu_7$         & $\Sigma^+-\Sigma^+$       &  702 --  746  &  $P(1-84)$, $R(1-86)$               &  236 &  0.5 \\[-0.5ex]
  $\nu_4$               & $\Sigma^+-\Sigma^+$       &  845 --  877  &  $P(5-52)$, $R(2-49)$               &   89 &  1.0 \\[-0.5ex]
  $\nu_5+\nu_7$         & $\Sigma^+-\Sigma^+$       &  868 --  909  &  $P(1-70)$, $R(0-64)$               &  131 &  1.0 \\[-0.5ex]
  $2\nu_6$              & $\Sigma^+-\Sigma^+$       &  998 -- 1030  &  $P(2-78)$, $R(1-76)$               &  153 &  0.5 \\[-0.5ex]
  $\nu_6+2\nu_7-\nu_7$  & $\Pi-\Pi$                 &  706 --  734  &  $P(6-45)$, $R(4-44)$               &  161 &  0.5 \\[0.5ex]
 \enddata
 \tablenotetext{a}{Estimated measurement accuracy.}
 \tablecomments{Asterisks label perturbation enhanced bands.}
\end{deluxetable*}

\subsection{Infrared spectrum}
\begin{figure}[t!]
  \centering
  \includegraphics[angle=0,width=0.81\textwidth]{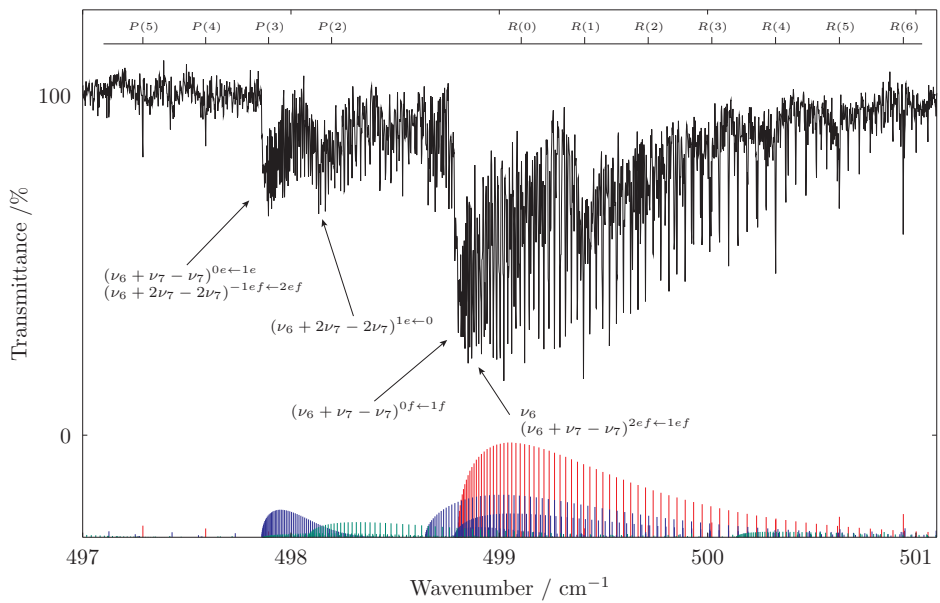}
  \caption{Portion of the infrared spectrum of HC$_3$N showing the region of the $\nu_6$ 
           band centre.
           The upper axis indicates $P,R$ line assignments for the fundamental $\nu_6$ band.
           The stick spectrum indicates the $\nu_6$ (red), the $\nu_6+\nu_7-\nu_7$
           (blue), and the $\nu_6+2\nu_7-2\nu_7$ (green) bands. 
           Line positions and relative intensities were calculated using the spectroscopic 
           constants of Tables~\ref{tab:rovib1}--\ref{tab:rovib4}.
           Recording conditions: $T=298$\,K, $P=16$\,Pa, $L_\text{path}=4$\,m, 440~scans, 
           unapodised resolution 0.004\,\wn\@.
          }
  \label{fig:nu6Q}
\end{figure}

\begin{figure}[t!]
  \centering
  \includegraphics[angle=0,width=0.84\textwidth]{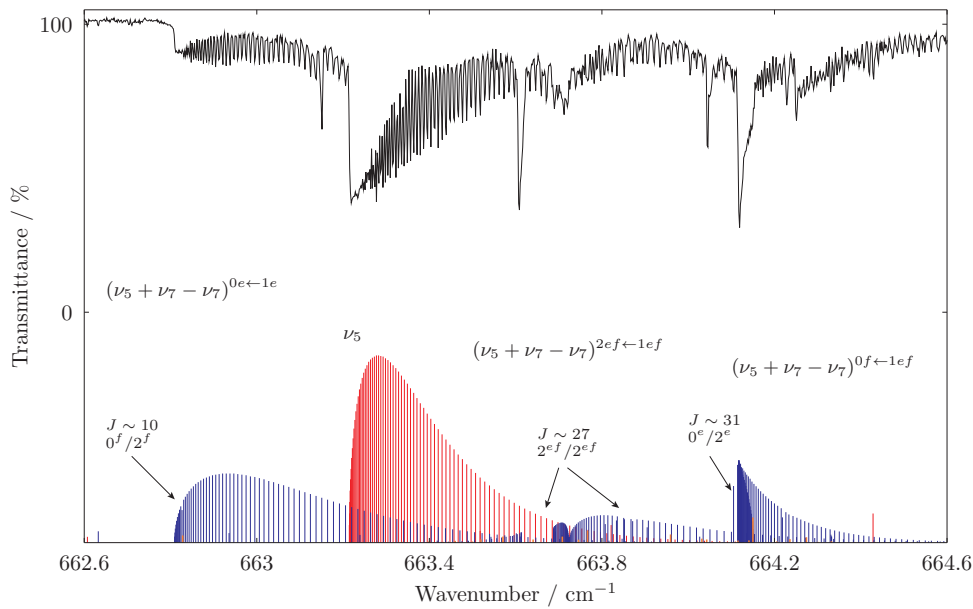}
  \caption{Portion of the infrared spectrum of HC$_3$N showing the region of the $\nu_5$ 
           band centre.
           The stick spectrum indicates the $\nu_5$ (red) and the $\nu_5+\nu_7-\nu_7$ 
           (blue) bands.
           Sparse orange sticks mark $4\nu_7-\nu_7$ lines whose intensity is enhanced by 
           the strong $\varv_7 = 4 \sim \varv_5 = \varv_7 = 1$ ro-vibrational mixing 
           (see text).
           Local perturbations due to the level avoided crossings are apparent in the 
           $Q$-branches and are indicated.
           The line positions and relative intensities were calculated using the 
           spectroscopic  constants of Tables~\ref{tab:rovib1}--\ref{tab:rovib4}.
           Recording conditions: $T=298$\,K, $P=16$\,Pa, $L_\text{path}=4$\,m, 440~scans, 
           unapodised resolution 0.004\,\wn\@.
          } 
  \label{fig:nu5Q}
\end{figure}

\indent\indent
The infrared spectra recorded in the laboratory cover the 450--1350\,\wn interval.
However, we decided to study in this work only ro-vibrational bands corresponding to 
vibrational levels with energy lower than 1000\,\wn.
This choice allowed us to perform a complete analysis of the low-lying vibrational states 
involved in anharmonic resonances present in \hctn.
In total, 13~IR~bands have been recorded and analysed.
Figure~\ref{fig:lev-scheme} shows the bottom part of the vibrational energy diagram of
\hctn up to \textit{ca.}~1300\,\wn.
The plot marks the investigated states and the IR transitions considered in the analysis.
The overview of the \hctn high-resolution vibrational spectrum over the full wavenumber
range covered by this study is shown in Figure~\ref{fig:overview}.
In the mid-IR region, the \hctn vibrational spectrum is dominated by the very strong 
bending fundamentals $\nu_5$ and $\nu_6$ located at $\sim 660$\,\wn and $\sim 500$\,\wn,
respectively.
Other weaker combination and overtone bands are visible at higher wavenumbers.
The $\nu_4$ stretching fundamental located at $\sim 873$\,\wn is very weak 
($I(\nu_5)/I(\nu_4)\sim 1/400$, \citealt{Jolly-JMS07-HC3N}) and could be observed only 
with long integration time (2100~scans) at a pressure of 270\,Pa and 4\,m optical 
path-length.
The spectral resolution was also lowered to 0.014\,\wn.
The strongest $\nu_5$ and $\nu_6$ band systems are particularly crowded because of the 
presence of numerous hot bands that are intense enough to be easily revealed.
Many $Q$-branches due to these fundamentals and their $\nu_7$-associated hot bands are 
clearly visible in the recorded spectrum near the corresponding band centres as shown 
in Figures~\ref{fig:nu6Q} and~\ref{fig:nu5Q}.

%%%% FIGURE NU6 Q-BRANCH

%%%% FIGURE NU5 Q-BRANCH

Table~\ref{tab:summIR} summarises the subset of bands that have been assigned and 
analysed in this study.
It comprises all but three bands that had been previously recorded by 
\citet{Arie-JMS90-HC3N}. 
The exceptions are the hot bands involving $2\nu_5-\nu_5$, $\nu_5+2\nu_7-2\nu_7$, and 
$\nu_5+\nu_6-\nu_6$, which lie at energies above 1100\,\wn, and they are all part 
of a complex network of resonances \citep{Mbosei-JMSt00-HC3N} that are currently under a 
separate investigation in greater detail.
Nevertheless, our chosen cut-off enables a complete and self-consistent analysis on the 
bottom part of the vibrational energy manifold of \hctn.
We would also like to point out that we have identified three new bands which have not
yet been reported on: the $\nu_4$ stretching fundamental and the 
perturbation enhanced $\nu_4 - \nu_7$ and $4\nu_7 - \nu_7$ bands that gain intensity from 
fairly strong interactions among the energy levels $\varv_4 = 1$, $\varv_5 = \varv_7 = 1$, 
and $\varv_7 = 4$.
These levels are all around 880\,\wn and are connected by the purely vibrational resonance
terms $\oH_{30}$, $\oH_{40}$, and $\oH_{50}$\@.
Local perturbations caused by avoided crossing among ro-vibrational levels are well 
visible in the infrared spectrum, particularly in the $Q$-branch of the $\nu_5+\nu_7-\nu_7$ 
hot band, as illustrated in Figure~\ref{fig:nu5Q}.

%%%%% TABLE ROT DATA
\begin{deluxetable*}{l ccccc}[t!]
 \tablecaption{Rotational data of \hctn used in the analysis
               \label{tab:summMW}}
 \tabletypesize{\footnotesize}
 \tablehead{
  \colhead{State}             &
  \colhead{$|k|$}             &
  \colhead{$J$ range}         &
  \colhead{Freq. range [GHz]} &
  \colhead{No. of lines}      &
  \colhead{Reference}
 }
 \startdata
  ground state              & $0$               &  0 -- 117 &     9 -- 1\,070        &  76 & dZ71,C77,C91,Y95,M00,T00,$\ast$     \\[-0.5ex]
  $\varv_7 = 1$             & $1^{e,f}$         &  2 -- 112 & 0.039 -- 1\,038\tnm{a} & 111 & L68,M78,dL85,Y86,C91,M00,T00,$\ast$ \\[-0.5ex]
  $\varv_7 = 2$             & $0,2^{e,f}$       &  0 -- 100 &     9 -- 923           & 124 & M78,Y86,M00,T00,$\ast$              \\[-0.5ex]
  $\varv_7 = 3$             & $(1,3)^{e,f}$     &  2 -- 100 &    27 -- 925           & 146 & L68,M78,Y86,M00,T00,$\ast$          \\[-0.5ex]
  $\varv_7 = 4$             & $0,(2,4)^{e,f}$   &  0 -- 100 &     9 -- 927           & 202 & M78,Y86,$\ast$                      \\[-0.5ex]
  $\varv_6 = 1$             & $1^{e,f}$         &  2 -- 100 & 0.021 -- 918\tnm{a}    & 105 & M78,dL85,Y86,M00,T00,Mor,$\ast$     \\[-0.5ex]
  $\varv_6 = 2$             & $0,2^{e,f}$       &  0 -- 99  &     9 -- 911           &  80 & M78,Y86,M00,$\ast$                  \\[-0.5ex]
  $\varv_5 = 1$             & $1^{e,f}$         &  2 -- 100 & 0.015 -- 917\tnm{a}    &  79 & dL85,M78,Y86,M00,T00,$\ast$         \\[-0.5ex]
  $\varv_4 = 1$             & $0$               &  0 -- 98  &     9 -- 897           &  39 & M78,Y86,M00,T00,$\ast$              \\[-0.5ex]
  $\varv_6 = \varv_7 = 1$   & $(0,2)^{e,f}$     &  0 -- 100 &     9 -- 922           & 126 & Y86,M00,T00,$\ast$                  \\[-0.5ex]
  $\varv_5 = \varv_7 = 1$   & $(0,2)^{e,f}$     &  0 -- 100 &     9 -- 920           & 211 & Y86,M00,$\ast$                      \\[-0.5ex]
  $\varv_6 = 1,\varv_7 = 2$ & $(-1,1,3)^{e,f}$  &  7 -- 99  &    73 -- 914           & 138 & Y86,M00,$\ast$                      \\[-0.5ex]
  \emph{interstate} &                           &  9 -- 56  &    92 -- 522           &  28 & $\ast$                              \\[0.5ex]
 \enddata
 \tablenotetext{a}{Includes MBER measurements from~\citeauthor{DeLeon-JCP85-HC3N} (Ref.~\cite{DeLeon-JCP85-HC3N}).}
 \tablecomments{dZ71 = \citet{deZafra-ApJ71-HC3N}, C77 = \citet{Cresw-JMS77-HC3N}, C91 = \citet{Chen-IJMW91-HC3N},
                Y95 = \citet{Yamada-ZN95-HC3N}, M00 = \citet{Mbosei-JMSt00-HC3N}, T00 = \citet{Thorw-JMS00-HC3N},
                M78 = \citet{Mall-MP78-HC3N}, Y86 = \citet{Yamada-JMS86-HC3N}, L68 = \citet{Laff-JMS68-HC3N},
                dL85 = \citet{DeLeon-JCP85-HC3N}, Mor = \citet{Mora-PhD}.
                Asterisk indicates that lines from this work are also included.}
\end{deluxetable*}

%%%% FIGURE ROTATIONAL SPECTRUM
\begin{figure*}[h!]
  \centering
  \includegraphics[angle=0,width=0.9\textwidth]{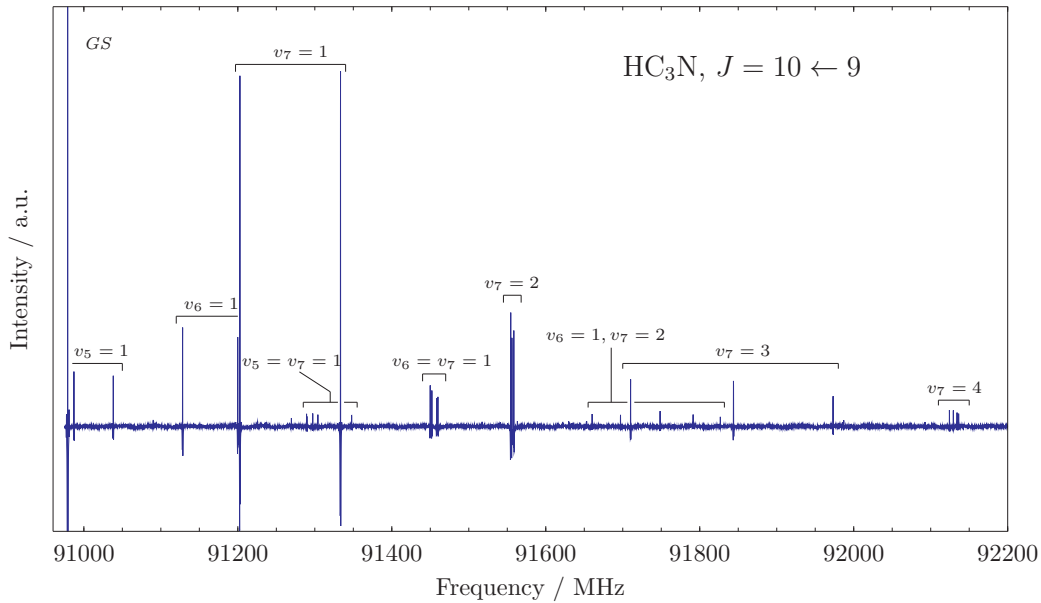}
  \caption{1.2\,GHz-long recording of the pure rotational spectrum of \hctn at 3\,mm
           wavelength in the region of the $J = 10\leftarrow 9$ transition.
           The out-of-scale feature at the left is the ground state line located at
           90979\,MHz.
           Line multiplets of 9 bending excited states are visible, they are:
           the fundamentals $\varv_5 = 1$, $\varv_6 = 1$, $\varv_7 = 1$; the overtones
           $\varv_7 = 2$, $\varv_7 = 3$, and $\varv_7 = 4$; and the combinations 
           $\varv_5 = \varv_7 = 1$, $\varv_6 = \varv_7 = 1$, and $\varv_6 = 1,\varv_7 = 2$\@.
           Recording conditions: $T = 298$\,K, $P = 0.7$\,Pa, 
           scan~speed$ = 2.5$\,MHz\,s$^{-1}$, with $RC = 3$\,ms\@.
          }
  \label{fig:mw}
\end{figure*}

\subsection{Rotational spectrum}
\indent\indent
The rotational spectra of the ground and vibrationally excited states of \hctn have 
already been investigated by several authors (see \S~\ref{sec:intro}).
Table~\ref{tab:summMW} presents an overview of the rotational data used in the present 
ro-vibrational analysis with the corresponding references.
Besides previous literature data, we report here a few unpublished sub-mm measurements,
as well as a new set of lines recorded recently in Bologna and in Garching.
These new data sets have been acquired so as to work out ambiguities that arose by revising 
previous literature data (e.g., the sub-mm data reported in \citealt{Mbosei-JMSt00-HC3N} 
are often affected by large uncertainties) and to improve the frequency coverage of some 
coarsely sampled spectra.
The maximum effort has been applied to the study of the strongly interacting states
$\varv_5 = \varv_7 = 1$ and $\varv_7 = 4$, for which 300 new lines --- including 28 
perturbation-enhanced cross-ladder transitions --- have been recorded.

Figure~\ref{fig:mw} provides a hint of the complexity of the vibrationally excited 
spectrum of \hctn.
The 1.2\,GHz-long spectral scan has been recorded in the region of the $J=10\leftarrow 9$ 
rotational transition.
All the vibrational satellites extend to the high-frequency side with respect to the 
ground state line (out-of-scale in the plot) located at 90979\,MHz.
In this excerpt, the $l$-type resonance patterns of all the bending excited states 
treated in the global analysis are visible.

%%%%% TABLE FITTED LINES (EXCERPT)
\begin{deluxetable*}{r rrr c c
                     r rrr c @{\hskip 2em}
                     DDD cc}[!b]
 \tabletypesize{\footnotesize}
 \tablecaption{Measured line positions and least-squares residuals for \hctn
               \label{tab:fit}}
 \tablehead{
  \colhead{$J'$} & \colhead{$l'_5$} & \colhead{$l'_6$} & \colhead{$l'_7$} & \colhead{$k'$} & 
  \colhead{$\leftarrow$} &
  \colhead{$J$}  & \colhead{$l_5$}  & \colhead{$l_6$}  & \colhead{$l_7$}  & \colhead{$k$}  &
  \multicolumn{2}{c}{observed} & \multicolumn{2}{c}{residual} & \multicolumn{2}{c}{$\sigma^a$} & 
  \colhead{units} & \colhead{Ref.}
 }
 \decimals
 \startdata
 \mcl{1}{l}{\ldots} \\
 \mcl{4}{l}{$\varv_7=4$} \\
  8 & 0 & 0 &  0 & $0^e$ &  &   7 & 0 & 0 &  0 & $0^e$ &   73702.486  &  0.00009 & 0.020 & MHz & Y86 \\[-1ex]
  9 & 0 & 0 &  0 & $0^e$ &  &   8 & 0 & 0 &  0 & $0^e$ &   82913.690  & -0.00521 & 0.015 & MHz & TW  \\[-1ex]
 10 & 0 & 0 &  0 & $0^e$ &  &   9 & 0 & 0 &  0 & $0^e$ &   92124.342  &  0.00161 & 0.015 & MHz & TW  \\[-1ex]
 11 & 0 & 0 &  0 & $0^e$ &  &  10 & 0 & 0 &  0 & $0^e$ &  101334.352  & -0.00760 & 0.015 & MHz & TW  \\[-1ex]
 12 & 0 & 0 &  0 & $0^e$ &  &  11 & 0 & 0 &  0 & $0^e$ &  110543.700  &  0.00857 & 0.020 & MHz & Y86 \\[-1ex]
 \mcl{4}{l}{\ldots} \\
 \mcl{4}{l}{$\nu_5+\nu_7-\nu_7$} \\
 20 & 1 & 0 &  1 & $2^e$ &  &  21 & 0 & 0 &  1 & $1^e$ &  657.39883   & 0.00006  & 0.001 & \wn & TW \\[-1ex]
 21 & 1 & 0 &  1 & $2^e$ &  &  22 & 0 & 0 &  1 & $1^e$ &  657.10124   & 0.00029  & 0.001 & \wn & TW \\[-1ex]
 22 & 1 & 0 &  1 & $2^e$ &  &  23 & 0 & 0 &  1 & $1^e$ &  656.80393   & 0.00011  & 0.001 & \wn & TW \\[-1ex]
 23 & 1 & 0 &  1 & $2^e$ &  &  24 & 0 & 0 &  1 & $1^e$ &  656.50797   & 0.00016  & 0.001 & \wn & TW \\[-1ex]
 24 & 1 & 0 &  1 & $2^e$ &  &  25 & 0 & 0 &  1 & $1^e$ &  656.21416   & 0.00019  & 0.001 & \wn & TW \\[-1ex]
 25 & 1 & 0 &  1 & $2^e$ &  &  26 & 0 & 0 &  1 & $1^e$ &  655.92520   & 0.00025  & 0.001 & \wn & TW \\[-1ex]
 \mcl{4}{l}{\ldots} \\
 \enddata
 \tablenotetext{a}{Assumed uncertainty for statistical weight calculation (see text).}
 \tablecomments{Y86 = \citet{Yamada-JMS86-HC3N}, TW = this work, \ldots}
\end{deluxetable*}

\section{Analysis} \label{sec:anal}
\indent\indent
The sample of IR and pure rotational data contains about 3400 ro-vibrational lines
for $13$ bands plus some 1500 pure rotational lines for 12~vibrational states.
The latter measurements extend over a very broad frequency interval that ranges from
the radio frequencies to the THz regime.
The composition and the general features of the data set are summarised in 
Tables~\ref{tab:summIR} and~\ref{tab:summMW}.
A different weighting factor, $w_i = 1/\sigma_i^2$ has been given to each $i$-th datum 
in order to take into account the different measurement precisions $\sigma$\@.

Two different uncertainties have been used for the present IR measurement: 
$\sigma = 0.5\times 10^{-3}$\,\wn for most measurements, whereas 
$\sigma = 1\times 10^{-3}$\,\wn has been adopted for the very weak $\nu_4$, and for a few 
other bands more affected by anharmonic resonances.
A summary of the weighting scheme is reported in Table~\ref{tab:summIR}.
For pure rotational lines, we adopted the general rule of retaining the weights used
in the original works.
In the cases when this information was missing, we adopted the ones provided by
\citet{Yamada-ZN95-HC3N} and \citet{Thorw-JMS00-HC3N}, who performed spectral analyses 
including data from the literature.
As far as our new millimetre/sub-millimetre measurements are concerned, we typically 
adopted an uncertainty of 15\,kHz\@. 
When necessary, a suitable different $\sigma$ was used.

The spectral analysis has been performed using a custom \textsc{Python} code which uses 
the \textsc{SPFIT} program \citep{Pick-JMS91-calpgm} as the computational core.
All the $l$-sublevels of the $12$ vibrational states are simultaneously represented in 
a $37\times 37$ ro-vibrational energy matrix, that is set up for each $J$ using the 
Hamiltonian described in \S~\ref{sec:theor}. 
This matrix is then reduced to a block-diagonal form and each block is separately
diagonalised to give the energy eigenvalues which are then compared to the observed
ro-vibrational terms.
The coefficients of the molecular Hamiltonian are optimised through an iterative
least-squares procedure, which delivers effective spectroscopic constants for
each individual state [Eqs.~\eqref{eq:diag}--\eqref{eq:dk4b}] plus a set of resonance
parameters [Eqs.~\eqref{eq:res30-466}--\eqref{eq:res42b}].

Actually, not all the coefficients of the Hamiltonian terms that matter for a given
vibrational state could be determined from the available experimental data.
For example, in states with $l_t = 1$, the $l$-dependent contributions expressed by the
parameters $x_{L(tt)}$, $d_{JL(tt)}$, and $h_{JL(tt)}$ merely produce additive terms
to the corresponding $G_v$, $B_v$, and $D_v$ constants [see Eq.~\eqref{eq:diag}] and
cannot be adjusted in the fit.
A few other adjustable constants turned out to be ill-determined due to the correlations
and were thus held fixed in the analysis.
Reliable constraints for these spectroscopic parameters have been obtained from the 
present experimental analysis: no assumptions derived from related molecules or 
theoretical calculations have been adopted.
The fixed spectroscopic constants of a given vibrational level have been derived from 
the corresponding optimised values obtained for other levels belonging to the same 
vibrational manifold considering, whenever feasible, a linear $\varv$ dependence.
The fitting procedure has thus been repeated until convergence of these 
inter-/extrapolated values was achieved.

The list of the analysed transitions frequencies/wavenumbers, including the least-squares 
residuals and the estimated measurement uncertainties is provided as digital supporting data.
An excerpt is presented here in Table~\ref{tab:fit} for guidance.
The data will also be available in the Cologne Database for Molecular Spectroscopy
\citep{Endres-JMS16-CDMS}
\footnote{http://www.astro.uni-koeln.de/site/vorhersagen/daten/HCCCN/vibs-up-to-1000cm-1}.
The spectroscopic parameters resulting from the global fit procedure are gathered in 
Tables~\ref{tab:rovib1}--\ref{tab:rovib4}.
Some details concerning the analysis are given in the following subsections.

%%%%% TABLE FIT RESULTS: SINGLY EXCITED
\begin{deluxetable*}{ll DDDDD}[t!]
 \tabletypesize{\footnotesize}
 \tablecaption{Results of the ro-vibrational analysis performed for \hctn: ground and singly-excited states
               \label{tab:rovib1}}
 \tablehead{
    \colhead{parameter}        &
    \colhead{units}            &
    \twocolhead{GS}            &
    \twocolhead{$\varv_4 = 1$} &
    \twocolhead{$\varv_5 = 1$} &
    \twocolhead{$\varv_6 = 1$} &
    \twocolhead{$\varv_7 = 1$}
 }
 \decimals
 \startdata
 $G_v$        &  \wn    &     .            &  878.312(17)    &  663.368484(31)  &  498.733806(26)  &  221.838739(33)  \\[-0.5ex]
 $x_{L(tt)}$  &  GHz    &     .            &     .           &    0.0^a         &    6.59^b        &   21.7972^b      \\[-0.5ex]
 $y_{L(tt)}$  &  MHz    &     .            &     .           &    0.0^a         &    0.0^a         &   -2.10^b        \\[-0.5ex]
 $B_v$        &  MHz    & 4549.058614(30)  & 4538.0977(21)   & 4550.62412(17)   &  4558.301481(60) & 4563.525640(61)  \\[-0.5ex]
 $D_v$        &  kHz    &    0.5442578(96) &    0.545383(83) &    0.545852(25)  &    0.554436(10)  &    0.568004(10)  \\[-0.5ex]
 $H_v$        &  mHz    &    0.0509(12)    &    0.0378(20)   &    0.0509(12)    &    0.06450(57)   &    0.10468(52)   \\[-0.5ex]
 $L_v$        &  nHz    &   -0.329(42)     &   -0.329^b      &   -0.329^b       &   -0.329^b       &   -0.329^b       \\[-0.5ex]
 $d_{JL(tt)}$ &  kHz    &     .            &     .           &    0.0^a         &   12.75^b        &  -12.287^b       \\[-0.5ex]
 $h_{JL(tt)}$ &  Hz     &     .            &     .           &    0.0^a         &    0.0^a         &    0.2443^b      \\[-0.5ex]
 $l_{JL(tt)}$ & $\mu$Hz &     .            &     .           &    0.0^a         &    0.0^a         &   -0.274^b       \\[-0.5ex]
 $q_t$        &  MHz    &     .            &     .           &    2.53870(11)   &    3.5821947(42) &    6.5386444(58) \\[-0.5ex]
 $q_{tJ}$     &  Hz     &     .            &     .           &   -1.3382(75)    &   -2.0611(20)    &  -16.2870(43)    \\[-0.5ex]
 $q_{tJJ}$    &  mHz    &     .            &     .           &    0.0^a         &    0.0^a         &   56.98(33)      \\[0.5ex]
 \enddata
 \tablenotetext{a}{Constrained.}
 \tablenotetext{b}{Assumed value, held fixed in the fit (see text).}
 \tablecomments{The numbers in parentheses are 1$\sigma$ uncertainties expressed in units of the last quoted digit.}
\end{deluxetable*}

%%%%% TABLE FIT RESULTS: MULTIPLY EXCITED
\begin{deluxetable*}{ll DDDD}[h!]
 \tabletypesize{\footnotesize}
 \tablecaption{Results of the ro-vibrational analysis performed for \hctn: 
               overtone bending states
               \label{tab:rovib2}}
 \tablehead{
    \colhead{parameter}        &
    \colhead{units}            &
    \twocolhead{$\varv_6 = 2$} &
    \twocolhead{$\varv_7 = 2$} &
    \twocolhead{$\varv_7 = 3$} &
    \twocolhead{$\varv_7 = 4$}
 }
 \decimals
 \startdata
 $G_v$        & \wn     &  997.913(17)    &  442.899036(61)  &  663.2205(29)   &  882.85147(21)  \\[-0.5ex]
 $x_{L(tt)}$  & GHz     &    6.59(13)     &   21.62866(55)   &   21.4398(12)   &   21.2814(15)   \\[-0.5ex]
 $y_{L(tt)}$  & MHz     &    0.0^a        &   -2.10^b        &   -2.10^b       &   -2.100(75)    \\[-0.5ex]
 $B_v$        & MHz     & 4567.4528(17)   & 4577.966834(78)  & 4592.38340(20)  & 4606.77431(27)  \\[-0.5ex]
 $D_v$        & kHz     &    0.564556(14) &    0.5922476(90) &    0.617092(43) &    0.642476(16) \\[-0.5ex]
 $H_v$        & mHz     &    0.07607(75)  &    0.15789(43)   &    0.2103(16)   &    0.27016(88)  \\[-0.5ex]
 $L_v$        & nHz     &   -0.329^b      &   -0.329^b       &   -0.329^b      &   -0.329^b      \\[-0.5ex]
 $d_{JL(tt)}$ & kHz     &   12.75(41)     &  -13.116(23)     &  -14.002(26)    &  -14.803(17)    \\[-0.5ex]
 $h_{JL(tt)}$ & Hz      &    0.0^a        &    0.2035(16)    &    0.1433(72)   &    0.1122(16)   \\[-0.5ex]
 $l_{JL(tt)}$ & $\mu$Hz &    0.0^a        &   -1.261^b       &   -2.25(24)     &   -3.235(96)    \\[-0.5ex]
 $q_t$        & MHz     &    3.582195^b   &    6.563988(58)  &    6.587477(36) &    6.612440(40) \\[-0.5ex]
 $q_{tJ}$     & Hz      &   -2.0611^b     &  -16.5793^b      &  -16.872(10)    &  -17.3008(98)   \\[-0.5ex]
 $q_{tJJ}$    & mHz     &    0.0^a        &   54.31^b        &   51.64(64)     &   56.96(56)     \\[-0.5ex]
 $u_{tt}$     & Hz      &     .           &   -0.1829^b      &   -0.1506(42)   &   -0.11825(50)  \\[0.5ex]
 \enddata
 \tablenotetext{a}{Constrained.}
 \tablenotetext{b}{Assumed value, held fixed in the fit (see text).}
 \tablecomments{The numbers in parentheses are 1$\sigma$ uncertainties expressed in units of the last quoted digit.}
\end{deluxetable*}

%%%%% TABLE FIT RESULTS: COMBINATIONS
\begin{deluxetable*}{ll DDD}[b!]
 \tabletypesize{\footnotesize}
 \tablecaption{Results of the ro-vibrational analysis performed for \hctn:  bend--bend 
               combination states 
               \label{tab:rovib3}}
 \tablehead{
    \colhead{parameter}        &
    \colhead{units}            &
    \twocolhead{$\varv_5 = \varv_7 = 1$} &
    \twocolhead{$\varv_6 = \varv_7 = 1$} &
    \twocolhead{$\varv_6 = 1, \varv_7 = 2$}
 }
 \decimals
 \startdata
 $G_v$          & \wn     &  885.37215(63)  &  720.293173(30)  &    941.070371(59)  \\[-0.5ex]
 $x_{L(tt)}$    & GHz     &    0.0^a        &    6.59^b        &      6.59^b        \\[-0.5ex]
 $x_{L(77)}$    & GHz     &   21.7972^b     &   21.7256^b      &     21.6541(11)    \\[-0.5ex]
 $x_{L(t7)}$    & GHz     &   19.277(20)    &   17.12595(41)   &     17.12142(89)   \\[-0.5ex]
 $y_{L(77)}$    & MHz     &   -2.10^b       &   -2.10^b        &     -2.10^b        \\[-0.5ex]
 $r_{t7}$       & GHz     &    6.999(39)    &  -11.77173(62)   &    -11.4965(12)    \\[-0.5ex]
 $r_{t7J}$      & kHz     &   -0.0252(14)   &  -12.670(74)     &    -10.205(90)     \\[-0.5ex]
 $r_{t7JJ}$     & Hz      &    1.117(88)    &    2.029(96)     &      2.047(23)     \\[-0.5ex]
 $B_v$          & MHz     & 4565.08511(80)  & 4572.861121(64)  &   4587.39057(14)   \\[-0.5ex]
 $D_v$          & kHz     &    0.569558(21) &    0.578017(11)  &      0.6021526(97) \\[-0.5ex]
 $H_v$          & mHz     &    0.10346(64)  &    0.12639(62)   &      0.18979(54)   \\[-0.5ex]
 $L_v$          & nHz     &   -0.329^b      &   -0.329^b       &   -0.329^b         \\[-0.5ex]
 $d_{JL(tt)}$   & kHz     &    0.0^a        &   12.75^b        &     12.75^b        \\[-0.5ex]
 $d_{JL(77)}$   & kHz     &  -12.287^b      &   -9.408^b       &     -6.529(50)     \\[-0.5ex]
 $h_{JL(77)}$   & Hz      &    0.2443^b     &    0.2443^b      &      0.2035^b      \\[-0.5ex]
 $l_{JL(77)}$   & $\mu$Hz &   -0.274^b      &   -0.274^b       &     -1.261^b       \\[-0.5ex]
 $d_{JL(t7)}$   & kHz     &  -21.22(61)     &   55.700(84)     &     50.580(42)     \\[-0.5ex]
 $h_{JL(t7)}$   & Hz      &    1.03(10)     &    0.0^a         &      0.0^a         \\[-0.5ex]
 $q_t$          & MHz     &    2.56538(23)  &    3.62324(50)   &      3.67072(44)   \\[-0.5ex]
 $q_{tJ}$       & Hz      &   -1.4356(88)   &   -2.2429^b      &     -2.426(10)     \\[-0.5ex]
 $q_7$          & MHz     &    6.53864^b    &    6.59341(21)   &      6.61796(19)   \\[-0.5ex]
 $q_{7J}$       & Hz      &  -16.287^b      &  -16.386(11)     &    -16.6165(70)    \\[-0.5ex]
 $q_{7JJ}$      & mHz     &   57.0^b        &   57.0^b         &     54.3^b         \\[-0.5ex]
 $u_{77}$       & Hz      &     .           &     .            &     -0.1854(36)    \\[-0.5ex]
 $u_{t7}$       & Hz      &   -1.14(11)     &   -2.211(98)     &     -2.259(24)     \\[-0.5ex]
 $q_{677}$      & kHz     &     .           &     .            &    -14.548(99)     \\[0.5ex]
 \enddata
 \tablenotetext{a}{Constrained.}
 \tablenotetext{b}{Assumed value, held fixed in the fit (see text).}
 \tablecomments{The numbers in parentheses are 1$\sigma$ uncertainties expressed in units 
                of the last quoted digit.}
\end{deluxetable*}

\subsection{Isolated states} \label{sec:anal:iso}
\indent\indent
Some of the bands considered in the analysis do not show any evidence of perturbation,
the involved vibrational states have thus been considered as isolated.
They are: the ground state, the $\varv_6 = 1$ and $\varv_7 = 1$, the $\varv_7 = 2$, 
and the $\varv_6 = \varv_7 = 1$ and $\varv_6 = 1,\varv_7 = 2$ bend--bend combination 
states.
Experimental information about these excited states derive from the measurements of 
several IR bands, namely: $\nu_6$, $\nu_6+\nu_7$, $\nu_6+\nu_7-\nu_7$, 
$\nu_6+2\nu_7-\nu_7$, and $\nu_6+2\nu_7-2\nu_7$ (see Table~\ref{tab:summIR}).
This rich set of data makes it possible to derive accurate vibrational energies for all 
the states, including the $\varv_7=1$ and $\varv_7=2$ bending levels, even if the $\nu_7$ 
fundamental and $2\nu_7$ overtone bands were not directly observed.

The pure rotational data available for these vibrational states include metre-wave 
molecular beam electric resonance measurements (MBER), direct $l$-type measurements at 
centimetre wavelength, and rotational spectra up to the THz regime 
(see Table~\ref{tab:summMW} for the bibliographic references).
For the ground state we recorded a few high-$J$ lines located above 1\,THz in order to 
improve the determination of the quartic ($D_0$) and sextic ($H_0$) centrifugal distortion 
constants, and a subset of very precise ($\sigma = 2$\,kHz) measurements at 3\,mm\@.
For the $\varv_6 = 1,\varv_7 = 2$ bend--bend combination state, less extensive rotational 
data were available.
We thus carried out new measurements in the 270--700\,GHz frequency range in order to 
achieve a satisfactory $J$ sampling to accurately model the $l$-type resonance effects.

Highly precise values for the rotational ($B_v$) and the quartic ($D_v$) centrifugal 
distortion constants have been obtained for all the states.
The derived $1\sigma$ standard errors of the $B_v$ parameters are a few tens of Hz 
(140\,Hz for $\varv_6 = 1,\varv_7 = 2$), while those of the $D_v$ are of the order of 
0.1\,mHz, or better.
We obtained a well determined (12\%) estimate for the octic centrifugal distortion 
constant ($L_v$) for the ground, and the sextic constants ($H_v$) have been determined 
for all the states with good precision (0.5--5\%). 
These values show a very smooth linear dependence on the vibrational quantum numbers 
$\varv_6$ and $\varv_7$.

Even vibrational trends are also observed for some high-order $l$-type parameters, 
such as $q_J$, $r_J$, $u_{67}$.
Anomalies in the fitted values of these small coefficients are very sensitive indicators 
of spectral perturbations, hence the observed regular behaviour allow us to 
rule out various interaction terms.
They are: the Coriolis-type $\oH_{31}$, which accounts for the 
$\varv_6 = 1$ $\sim$ $\varv_7 = 2$ and $\varv_5 = 1$ $\sim$ $\varv_6 = \varv_7 = 1$ 
couplings; and the anharmonic $\oH_{40}$ producing the 
$\varv_5 = 2$ $\sim$ $\varv_6 = \varv_7 = 1$ interaction.
These resonances had been considered by \citet{Fayt-JMSt04-HC3N15,Fayt-CP08-HC3N15} in the 
global analysis of the HCCC$^{15}$N but they proved to be negligible for \hctn in the 
$J$ range sampled in the present investigation.

\subsection{The interacting states $\varv_5=1$ $\sim$ $\varv_7=3$} \label{sec:anal:5-37}
\indent\indent
The $\varv_5=1$ is the highest-energy bending fundamental (\mbox{H$-$C$\equiv$C}) and is
located $\sim$ 663\,\wn above the ground state.
This state has a $\Pi$ symmetry and can interact with the $l_7 = 1$ level ($\Pi$ symmetry) 
of the nearby $\varv_7 = 3$ manifold.
The vibrational harmonic energy difference between these two states is 
$\omega_5 - 3\omega_7\simeq$ $-17.9$\,\wn (Pietropolli Charmet, in prep.).
The anharmonic value, which can be obtained from our experimental data, is 
$G_{\varv_5} - G_{3\varv_7} - x_{L(77)}\simeq$ $-0.57$\,\wn\@.
The two states are actually closely degenerate and can be coupled by the $\oH_{40}$ 
term of the effective Hamiltonian whose matrix elements are given in 
Eq.~\eqref{eq:res40-5777}.
Both $e$ and $f$ parity sublevels are affected with contributions of similar magnitude.
Due to the small value of the $C_{40}$ quartic coefficient, the resonance is weak: 
at low values of the $J$ quantum number, where the effects are stronger, each ro-vibrational 
level is pushed away by $\approx$ 60\,MHz\@.
The resonance strength decreases with increasing $J$ because the upper level, 
$\varv_7 = 3$, has a higher value of the rotational constant $B_v$, and therefore the 
two interacting states become more and more separate in energy with the rotational 
excitation.

Experimental information about these interacting states are provided by the $\nu_5$ band 
recorded in the IR (see Table~\ref{tab:summIR}) and by the pure rotational spectra for the
vibrationally excited $\varv_5 = 1$ and $\varv_7 = 3$ states.
The ro-vibrational data allow an accurate determination of the vibrational energy 
of the $\varv_5 = 1$ state, while the analysis of the resonance provides the absolute 
position of the $\varv_7 = 3$ level which is determined with a standard error of 
$3\times 10^{-3}$\,\wn\@.

The present set of data are not sensitive to the value of the $x_{L(55)}$ anharmonicity
constant that merely acts as an addictive term to the $\varv_5 = 1$ vibrational energy.
Its value has been fixed to zero and its contribution is thus included in the 
corresponding $G_v$\@.
The same applies to the $d_{JL(55)}$, $h_{JL(55)}$, and $l_{JL(55)}$ coefficients 
expressing the $l$ dependency of the rotational ($B_v$), quartic ($D_v$), and sextic ($H_v$) 
centrifugal distortion constants.

For $\varv_7 = 3$, we adjusted all the Hamiltonian coefficients given in 
Eqs.~\eqref{eq:diag} and~\eqref{eq:dk2a}, except $y_{L(77)}$, which was held fixed at the 
value derived for the $\varv_7 = 4$ bending level (see \S~\ref{sec:anal:reson}).
For this state we also optimised the $|\Delta_k|=4$ parameter $u_{77}$ that models 
high-order $l$-type resonance effects between $l_7 = 1$ and $l_7 = 3$ sublevels.
Its value was determined with a 3\% standard uncertainty and its order of magnitude is
in agreement with what is expected (see \S~\ref{sec:disc}).
Precise determinations have also been obtained for $H_v$ (0.5\%) as well as other 
high-order parameters such as $q_{tJJ}$ (1\%), $h_{JL(77)}$ (4\%), and $l_{JL(77)}$ (7\%).

The resonance effects have been modelled by adjusting the main parameter $C_{40}^{(5777)}$
together with its $J$-dependence coefficient $C_{42}^{(5777J)}$, which has been determined 
with a precision of about 10\%.
The latter, small parameter is necessary to reproduce the rotational mixing between 
$\varv_5=1,l_5=1$ and $\varv_7=3,l_7=1$ sub-states and its trend over the fairly large $J$
interval (2--100) sampled by the present experimental data.

%%%%% FIGURE RESONANCE SCHEME
\begin{figure}[b!]
  \centering
  \includegraphics[angle=0,width=0.6\textwidth]{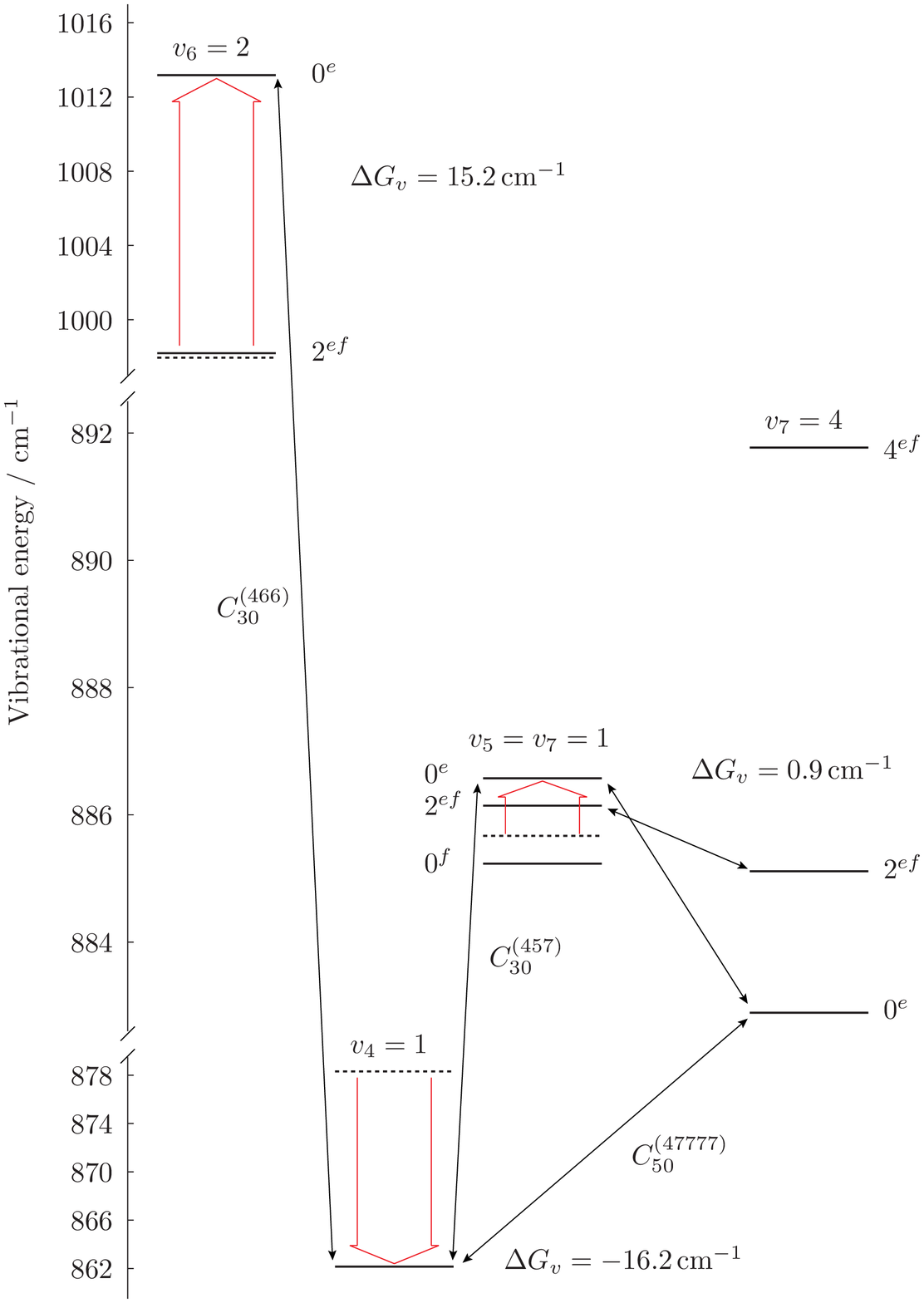}
  \caption{Vibrational energy level diagram for the interacting states $\varv_4 = 1$, 
           $\varv_5 = \varv_7 = 1$, $\varv_6 = 2$, and $\varv_7 = 4$ of \hctn\@.
           Thin arrows (black) indicate the main vibrational coupling taken into account. 
           Large arrows (red) illustrate the vibrational energy displacements produced 
           by the anharmonic resonances; the hypothetical unperturbed level positions are 
           plotted with dashed lines.
           The weak $\Delta k = \pm 2$ interactions produced by $\oH_{42}$ effective 
           Hamiltonian term between the $\varv_5 = \varv_7 = 1$ and $\varv_7 = 4$ 
           states are not indicated.
          }
  \label{fig:reso}
\end{figure}

%%%%% FIGURE FORTRAN DIAGRAM
\begin{figure*}[tbh]
  \centering
  \includegraphics[angle=0,width=0.8\textwidth]{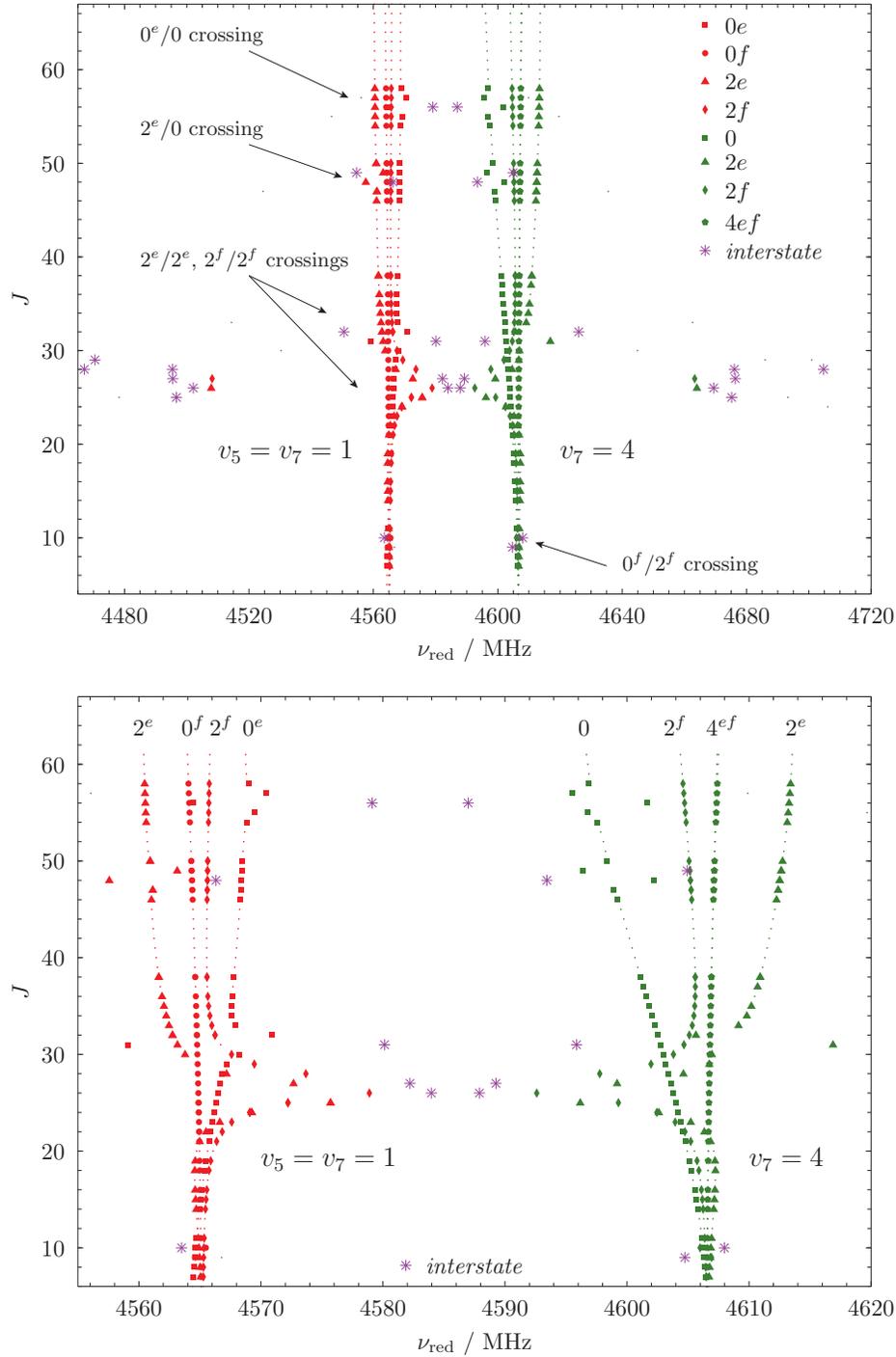}
  \caption{Reduced frequency diagram for the $\varv_5 = \varv_7 = 1$ (red symbols) and 
           $\varv_7 = 4$ (green symbols) interacting states of \hctn.
           The quantity plotted onto $x$-axis is 
           $\nu_\text{red} = [\nu + 4D(J+1)^3]/2(J+1)$\@.
           Solid symbols denote experimental values, whereas small dots indicate calculated
           values based on the parameters of Tables~\ref{tab:rovib1}--\ref{tab:rovib4}.
           The most pertubed transitions, in the proximity of the crossings, are labelled 
           using the method implemented in the SPFIT code 
           \citep{Pick-JMS91-calpgm}.
           The bottom panel shows a detail of the upper plot in the $\nu_\text{red}$ range
           from~4555 to~4620\,MHz and most of the interstate transitions are not visible 
           here.
          }
  \label{fig:fortrat}
\end{figure*}

\subsection{The resonance system $\varv_4=1$ $\sim$ $\varv_5=\varv_7=1$ $\sim$ 
                                 $\varv_6=2$ $\sim$ $\varv_7=4$}
\label{sec:anal:reson}
\indent\indent
In \hctn, there is a polyad of interacting states which pivots on the lowest energy
$\varv_4 = 1$ stretching fundamental located at $\sim$ 878\,\wn.
This state can be coupled with the $\Sigma$ ($l = 0$) sub-level of any nearby
doubly-excited bending state through cubic anharmonic resonances generated by
the $\oH_{30}$ term of the effective Hamiltonian.
In our analysis we have considered the interactions of $\varv_4 = 1$ with 
$\varv_5 = \varv_7 = 1$ and with $\varv_6 = 2$, which are located at 885\,\wn and 998\,\wn, 
respectively.
Furthermore, the $\varv_5 = \varv_7 = 1$ bend-bend combination state is coupled with the
$\varv_7 = 4$ level through the $\oH_{40}$ quartic anharmonic resonance.
This dyad is analogous to the one described in \S~\ref{sec:anal:5-37} and is obtained by 
adding one $\varv_7$ quantum to the fundamental dyad $\varv_5 = 1 \sim \varv_7 = 3$\@.
Finally, the weak quintic ($\oH_{50}$) resonance connecting $\varv_4 = 1$ and the $l = 0$
sub-level ($\Sigma$ symmetry) of the $\varv_7 = 4$ bending state has been also considered.

The detailed scheme of the energy level manifolds involved in this resonance system is
depicted in Figure~\ref{fig:reso}.
A major energy displacement ($\sim$ 16\,\wn) is experienced by the $\varv_4 = 1$ 
(pushed down) and by $\varv_6 = 2,l = 0$ substate (pushed up), because of the large value 
of the $\phi_{466}$ cubic force constant involved in the $\oH_{30}$ resonance term.
This effect is analogous to the Fermi resonance in triatomic molecules and is also present
in many linear polyatomic molecules, such as the \hctn isomer isocyanoacetylene
(HCCNC, \citealt{Vigou-JMS00-HCCNC}), the isoelectronic species diacetylene
(HC$_4$H, \citealt{Bizz-MP11-HC4H}) and the longer chain HC$_5$N \citep{Bizz-JMS05-HC5N}.
Though closer in energy to $\varv_4 = 1$, the $\varv_5 = \varv_7 = 1$ state is less 
affected by this resonance, because the corresponding $\phi_{457}$ cubic force constant is 
smaller.
Nonetheless, the resulting displacement of $\sim 1$\,\wn is enough to invert the relative
positions of the $k = 0^e$ and $k = 2^{ef}$ sub-levels, altering completely the $l$-type
resonance effects within the $\varv_5 = \varv_7 = 1$ manifold.

As shown in Figure~\ref{fig:reso}, the two interacting states $\varv_5 = \varv_7 = 1$ and 
$\varv_7 = 4$ are very close.
For $J = 2$, the energy difference is $\sim 3.7$\,\wn for $0^e$ and $\sim 1.1$\,\wn for 
$2^{ef}$ sub-levels.
These gaps decrease for increasing $J$, because the $\varv_7 = 4$ ro-vibrational levels
--- initially located below the corresponding $\varv_5 = \varv_7 = 1$ ones --- have a 
higher effective rotational constant 
($B_{\varv_7 = 4}$ - $B_{\varv_5 = \varv_7 = 1} \simeq 42$\,MHz).
The reduced frequency plot for these two states is presented in Figure~\ref{fig:fortrat}.
Deviations from linear behaviour are due to high-order effects: regular, slightly bent 
parabolas are produced by residual energy contributions depending on $J^6$ and produced 
by $l$-type resonances.
Abrupt changes in curvature and discontinuities are generated by avoided crossings 
between close degenerate levels.
These effects are very visible at $J\sim 26-27$ and $J\sim 56$, where almost exact
degeneracies occur between the $2^e/2^f$ and $0^e/0$ sublevels of the two states.
The lines show displacements as large as $\sim$ 3\,GHz with respect to their unperturbed
positions and, because of the strong ro-vibrational mixing, a series of cross-ladder
(interstate) transitions gain enough intensity to be readily detected.
These forbidden transitions are indicated by purple stars in Figure~\ref{fig:fortrat}.
Less striking perturbations also occur at $J\sim 9$ ($0^f/2^f$ crossing) and $J\sim 48$
($2^e/0$ crossing).

%%%%% TABLE FIT RESULTS: RESONANCE PARAMETERS
\begin{deluxetable*}{lll D}[b!]
 \tabletypesize{\footnotesize}
 \tablecaption{Results of the ro-vibrational analysis performed for \hctn: 
               resonance parameters 
               \label{tab:rovib4}}
 \tablehead{
  \colhead{interacting states} &
  \colhead{parameter}          &
  \colhead{units}              &
  \twocolhead{fitted value}
 }
 \decimals
 \startdata
 ($\varv_5 = 1$) $-$ ($\varv_7 = 3$)            & $C_{40}$   &  MHz &   784.4(42)       \\[-0.5ex]
                                                & $C_{42}^J$ &  kHz &   -29.1(37)       \\[-0.5ex]
 ($\varv_5 = \varv_7 = 1$) $-$ ($\varv_7 = 4$)  & $C_{40}$   &  MHz &   747.1(43)       \\[-0.5ex]
                                                & $C_{42}^J$ &  MHz &     0.1276(56)    \\[-0.5ex]
                                                & $C_{42a}$  &  kHz &    20.31(70)      \\[-0.5ex]
                                                & $C_{42b}$  &  kHz &     7.86(42)      \\[-0.5ex]
 ($\varv_4 = 1$) $-$ ($\varv_6 = 2$)            & $C_{30}$   &  \wn &    16.0275(81)    \\[-0.5ex]
                                                & $C_{32}^J$ &  MHz &    -0.5164(20)    \\[-0.5ex]
 ($\varv_4 = 1$) $-$ ($\varv_5 = \varv_7 = 1$)  & $C_{30}$   &  \wn &    -2.4161(34)    \\[-0.5ex]
 ($\varv_4 = 1$) $-$ ($\varv_7 = 4$)            & $C_{50}$   &  GHz &     3.458(24)     \\[-0.5ex]
                                                & $C_{52}^J$ &  kHz &    34.2(12)       \\[0.5ex]
 \enddata
 \tablecomments{The numbers in parentheses are 1$\sigma$ uncertainties expressed in 
                units of the last quoted digit.}
\end{deluxetable*}

For the states involved in this resonance system we have a great deal of experimental 
information.
In the IR, we recorded the $\nu_4$, $\nu_5 + \nu_7$, and $2\nu_6$ bands that provide the
energy position of the interacting levels.
Additional ro-vibrational information on all the $l$-sublevels come from the measurements 
of the $2\nu_6-\nu_6$ and from the $\nu_4-\nu_7$ and $4\nu_7-\nu_7$ difference bands
(see Table~\ref{tab:summIR}).
The latter are weak IR features whose intensity is enhanced by the ro-vibrational mixing 
produced by the $\oH_{30}$ and $\oH_{40}$ resonance terms.
An extensive pure rotational data set is also available for all four states given by 
previous and new measurements (see Table~\ref{tab:summMW}).
The data span a remarkable $J$ interval, from~0 to~100, covering the 9--920\,GHz frequency 
range.
In particular, we recorded~300 new lines for the pair of states 
$\varv_5 = \varv_7 = 1$/$\varv_7 = 4$, including~28 interstate transitions around the most 
perturbed $J$ values.
This wealth of data has enabled a complete analysis of the resonance 
system without using any assumption derived from theoretical calculations or extrapolated
from related molecules.
The absolute vibrational energy positions of all levels and most of the Hamiltonian 
coefficients have been adjusted in the least-squares fit.
In a few cases they have been held fixed to suitable values derived from related 
vibrational states.

The sextic centrifugal distortion constant ($H_v$) has been determined for all states 
with good precision (7\% at least).
The lines of the most perturbed $\varv_5 = \varv_7 = 1$ and $\varv_7 = 4$ states required 
a few additional high-order parameters in order to be fitted within experimental 
uncertainties.
Besides the $|\Delta k| = 4$ coefficients $u_{57}$ and $u_{77}$, for $\varv_7 = 4$ we had 
to adjust $y_{L(77)}$, which represents the $l^4$ dependence of the vibrational energy, and 
the high-order $l$-type doubling constant $q_{677}$ [see Eq.~\eqref{eq:dk2b}].
The resonance effects were accurately modelled by including the $J$-dependent coefficients
$C_{32}^{(457J)}$, $C_{32}^{(466J)}$, $C_{42}^{(5777J)}$, and $C_{52}^{(47777J)}$, plus 
the $|\Delta k| = 2$ parameters $C_{42a}^{(5777)}$, $C_{42b}^{(5777)}$\@.

The contribution of these terms is very small since their order of magnitude is between
$\kappa^3\omega_\text{vib}$ and $\kappa^5\omega_\text{vib}$, as described in 
Table~\ref{tab:odg}.
Nevertheless, their inclusion in the analysis is justified by the following considerations:
($i$) the wide $\varv$ and $J$ range sampled ($J_\text{max} = 100$, $\varv_\text{max} = 4$);
($ii$) the ``exact'' resonance occurring between $\varv_5 = \varv_7 = 1$ and $\varv_7 = 4$ 
ro-vibrational levels; ($iii$) the high precision frequency determination 
($\sigma = 15$\,kHz) of the most perturbed lines and of the interstate transitions, which 
are extremely sensitive to subtle resonance effects.

\section{Discussion} \label{sec:disc}
\indent\indent
In the present spectral analysis we have treated simultaneously the whole set of 
high-resolution data available for \hctn up to an energy of ca.\ 1000\,\wn\@.
Almost 5000 experimental transitions have been included in the least-squares fit: 
this resulted in the determination of 11~vibrational energies and 110~spectroscopic 
constants for 12~states.
The overall quality of the analysis is expressed by the weighted root-mean-square 
deviation, defined as
\begin{equation} \label{eq:rms}
 \sigma_\text{rms} = \sqrt{\frac{1}{N} \sum_{i=1}^N \left( \frac{\nu_i^\text{exp} 
                            - \nu_i^\text{calc\pha{p}}}{\sigma_i}\right)^2 } \,.
\end{equation}
We obtained $\sigma_\text{rms} = 0.875$, indicating that the experimental data set has 
been reproduced well within the estimated measurements accuracies.

The knowledge of the ro-vibrational spectrum of \hctn has been greatly improved.
The most important spectroscopic constants have been determined with very high precision:
$1\sigma$ uncertainties of ca.\ one part over $10^8$ have been obtained for the rotational
constant $B_v$ of most states, whereas the average precision of the quartic centrifugal 
distortion constant ($D_v$) and $l$-type doubling constant ($q_v$) is of a few parts over 
$10^5$ and $10^7$, respectively.
Several anharmonicity constants of the type $x_{L(tt)}$, $x_{L(tt')}$, and $r_{tt'}$, 
plus a number of high-order ro-vibrational parameters have been also determined with 
good precision.

Unlike some extensive studies published in the past 
\citep[e.g.,][]{Arie-JMS90-HC3N,Mbosei-JMSt00-HC3N}, we considered explicitly the $l$-type 
resonance effects among bending sublevels, thus obtaining a unique set of spectroscopic 
parameters for each vibrational state.
More importantly, by joining high-resolution IR data and pure rotational measurements, 
we attained a thorough modelling of the spectral perturbations produced by the anharmonic 
resonances in the bottom part of the vibrational energy manifold of \hctn.
Compared with the previous work of \citet{Yamada-JMS86-HC3N}, who performed a similar
treatment on a much more limited data set, we achieved a substantial improvement
in terms of completeness of the analysis and overall accuracy of the derived 
spectroscopic parameters.

It should be noted that \citet{Jolly-JMS07-HC3N} performed a global treatment of all the
data available for \hctn in \citeyear{Jolly-JMS07-HC3N}.
Their analysis was used to support accurate calculations of integrated IR intensity for 
the $\nu_5$ and $\nu_6$ hot band systems, but the complete results and the list of the 
derived spectroscopic parameters have not been published.
Our study is thus the first complete ro-vibrational analysis for \hctn presented so far:
the level of detail adopted in the description of the ro-vibrational energies and the 
wide $J$ interval spanned by the data set make our analysis particularly suited for
astrophysical applications.

\subsection{Molecular parameters} \label{sec:disc:gen}
\indent\indent
The methodology used for the present analysis implies the determination of one set of 
effective spectroscopic constants for each vibrational states, plus a few resonance 
parameters describing the various anharmonic couplings.
In this approach, the use of a ro-vibrational Hamiltonian that includes all the relevant 
interactions is critical to obtain state-specific parameters with clear physical meaning 
and reliable predictions for the unexplored spectral regions.
Indications on the validity of such a treatment can be derived by evaluating the 
$\varv$-dependence of the determined spectroscopic parameters.
For a given bending state-specific parameter, $a_\varv$, an empirical $\varv$-series 
expansion holds
\begin{equation} \label{eq:v-dep}
  a_\varv = a_e + \beta_a (\varv + 1) + \gamma_a (\varv + 1)^2 + \ldots \,, 
\end{equation}
where $a_e$ is the pseudo-equilibrium value of the $a$ constant purged from the specific 
$\varv$ dependence, while $\beta_a$ and $\gamma_a$ represent the expansion coefficients 
for the first and second order contributions, respectively.

From the results presented in Tables~\ref{tab:rovib1}--\ref{tab:rovib3}, the 
quantity $\beta_a$ can be evaluated for an extensive subset of spectroscopic constants 
upon excitation of the $\varv_6$ and $\varv_7$ quanta, showing that the state-specific 
parameters we determined for \hctn exhibit a remarkable regular behaviour.
Eq.~\eqref{eq:v-dep} applied to the most important spectroscopic parameters 
($B_v$, $D_v$, and $q_v$), shows a rapid convergence with only minor departures
from the linear trend: 0.5\% for rotational constant, ($B_v$), 3\% for the $q_7$ $l$-type 
doubling constant, 5\% and 10\% for quartic ($D_v$) and sextic ($H_v$) centrifugal 
distortion constant, respectively.
Even trends are also shown by the anharmonicity constants $x_{L(77)}$, $x_{L(67)}$, and 
by the $r_{67}$ vibrational $l$-type doubling parameter, whose converged values mildly 
decrease (1--2\%) upon $\varv_7$ excitation.
Furthermore, no obvious anomalies are exhibited by the high-order coefficients: maximum 
variations of $\sim$ 6\% are observed for $q_{7J}$, $\sim$ 10\% for $q_{7JJ}$ and 
$d_{JL(77)}$, $\sim$ 30\% for $u_{77}$ and $r_{67J}$, and $\sim$ 50\% for $u_{77}$ and 
$r_{67JJ}$\@.
These findings are very well comparable with the results of earlier global ro-vibrational 
analyses performed on related molecules \citep[e.g.,][]{Fayt-JMSt04-HC3N15}, and provide
a strong indication that the effective Hamiltonian adopted for \hctn is adequate
for the span and the precision of the available data set.

\subsection{Spectral predictions} \label{sec:disc:pred}
\indent\indent
With the spectroscopic constants presented in Tables~\ref{tab:rovib1}--\ref{tab:rovib4} 
we computed an extensive set of accurate ro-vibrational rest frequencies for all
the vibrational levels of \hctn below 1000\,\wn.
These spectral predictions are provided as digital supporting data, and consist of a 
compilation of IR wavenumbers for all the bands observed in this work 
(see Table~\ref{tab:summIR}), plus pure-rotational frequencies for all the states listed 
in Table~\ref{tab:summMW}, including inter-state transitions.
The $J$ interval selected for the calculation is $0-120$\@.
The quadrupole coupling due to the $^{14}$N nucleus has not been considered, thus the
computed frequencies for the $J=0-4$ pure rotational transitions correspond to the
hypothetical hyperfine-free, unsplit line positions.
The estimated uncertainty at the $1\sigma$ level of each transition frequency is determined 
statistically by the least-squares fits \citep{Albritton-1976}.
The data list will be also made available in the Cologne Database for Molecular Spectroscopy
\citep{Endres-JMS16-CDMS}.
An excerpt of the data listing is presented in Table~\ref{tab:pred} for guidance purposes:
the following columns are included:
\begin{description}
 \item[{\rm (1--5)}] $J'$, $l'_5$, $l'_6$, $l'_7$, $k'$.
                     Rotational, vibrational angular quantum numbers, 
                     and $e/f$ parity of the upper level.
 \item[{\rm (6--10)}] $J$, $l_5$, $l_6$, $l_7$, $k$.
                     Rotational, vibrational angular quantum numbers, 
                     and $e/f$ parity of the lower level.
 \item[{\rm (11)}] $\nu_{J'J}$. Predicted line position computed from the spectroscopic
                   constants of Tables~\ref{tab:rovib1}--\ref{tab:rovib4}.
 \item[{\rm (12)}] 1$\sigma$. Estimated error of the prediction at $1\sigma$ level.
 \item[{\rm (13)}] units. MHz or \wn. Applies to columns (11) and (12).
 \item[{\rm (14)}] $S_{J'J}$. H\"onl-London factor.
 \item[{\rm (15)}] $E_u/k$. Upper state energy in K.
 \item[{\rm (16)}] $g_u$. Upper state degeneracy.
\end{description}

%%%%% TABLE REST FREQUENCIES
\begin{deluxetable*}{r rrr c c
                  r rrr c @{\hskip 2em}
                  DD c DD r}[t!]
 \tabletypesize{\footnotesize}
 \tablecaption{Computed rest frequencies for \hctn
               \label{tab:pred}}
 \tablehead{
  \colhead{$J'$} & \colhead{$l'_5$} & \colhead{$l'_6$} & \colhead{$l'_7$} & \colhead{$k'$} & 
  \colhead{$\leftarrow$} &
  \colhead{$J$}  & \colhead{$l_5$}  & \colhead{$l_6$}  & \colhead{$l_7$}  & \colhead{$k$}  &
  \multicolumn{2}{c}{$\nu_{J'J}$} & \multicolumn{2}{c}{$1\sigma$} & \colhead{units} & 
  \multicolumn{2}{c}{$S_{J'J}$}   & \multicolumn{2}{c}{$E_u/k$}   & $g_u$
 }
 \decimals
 \startdata
 \mcl{4}{l}{\ldots} \\
 \mcl{4}{l}{$\varv_5 = \varv_7 = 1$} \\
   9 & 1 & 0 & -1 & $0^e$ &  &   8 & 1 & 0 & -1 & $0^e$ &  82159.6651  & 0.0016   & MHz  & 8.35  & 1295.3 & 19  \\[-1ex]
   9 & 1 & 0 &  1 & $2^e$ &  &   8 & 1 & 0 &  1 & $2^e$ &  82168.7926  & 0.0010   & MHz  & 8.48  & 1294.7 & 19  \\[-1ex]
   9 & 1 & 0 & -1 & $0^f$ &  &   8 & 1 & 0 & -1 & $0^f$ &  82170.2499  & 0.0014   & MHz  & 9.00  & 1293.4 & 19  \\[-1ex]
   9 & 1 & 0 &  1 & $2^f$ &  &   8 & 1 & 0 &  1 & $2^f$ &  82173.6651  & 0.0010   & MHz  & 8.48  & 1294.7 & 19  \\[-1ex]
 \mcl{4}{l}{\ldots} \\
 \mcl{4}{l}{$\nu_6$} \\
  18 & 0 & 1 &  0 & $1^e$ &  &  19 & 0 & 0 &  0 & $0^e$ &  493.12092   & 0.00003  & \wn  & 4.50  &  792.4 & 37 \\[-1ex]
  17 & 0 & 1 &  0 & $1^e$ &  &  18 & 0 & 0 &  0 & $0^e$ &  493.41536   & 0.00003  & \wn  & 4.25  &  784.6 & 35 \\[-1ex]
  16 & 0 & 1 &  0 & $1^e$ &  &  17 & 0 & 0 &  0 & $0^e$ &  493.71032   & 0.00003  & \wn  & 4.00  &  777.1 & 33 \\[-1ex]
  15 & 0 & 1 &  0 & $1^e$ &  &  16 & 0 & 0 &  0 & $0^e$ &  494.00578   & 0.00003  & \wn  & 3.75  &  770.1 & 31 \\[-1ex]
  14 & 0 & 1 &  0 & $1^e$ &  &  15 & 0 & 0 &  0 & $0^e$ &  494.30174   & 0.00003  & \wn  & 3.50  &  763.6 & 29 \\[-1ex]
  13 & 0 & 1 &  0 & $1^e$ &  &  14 & 0 & 0 &  0 & $0^e$ &  494.59820   & 0.00003  & \wn  & 3.25  &  757.5 & 27 \\[-1ex]
 \mcl{4}{l}{\ldots} \\
 \enddata
 \tablecomments{See column explanation in the text, \S~\ref{sec:disc:pred}.}\vspace{-3ex}
\end{deluxetable*}

The corresponding Einstein A-coefficients for spontaneous emission can then be calculated
for each $J'\rightarrow J$ line using \citep{Simec-JQRST07-HITRAN}
\begin{equation} \label{eq:A}
 A_{J'J} = \frac{16\pi^3}{3\epsilon_0hc^3}\frac{\nu^3_{J'J}}{g_u} S_{J'J}\mathfrak{R}^2 \,,
\end{equation}
where $\nu_{J'J}$ is the transition frequency, $S_{J'J}$ is the computed rotational line 
strength factor (H\"onl-London factor) as given in Table~\ref{tab:pred}, $g_u$ is the 
upper level degeneracy (also given in Table~\ref{tab:pred}), 
and $\mathfrak{R}^2$ is the squared transition dipole moment (units of C$^2\cdot$m$^2$).
For a pure rotational transition $\mathfrak{R}^2 = \mu^2$, where $\mu$ is the permament
electric dipole moment.
For a vibration-rotation transition the squared transition dipole moment is
\begin{equation} \label{eq:sqdip}
 \mathfrak{R}^2 = |R^0_{v'v}|^2F\,,
\end{equation}
where $|R^0_{v'v}|$ is the rotationless vibrational transition dipole moment including 
the appropriate vibrational factors for the hot bands \citep{Fayt-JMS04-NC4N}, 
and $F$ is the Herman--Wallis factor \citep{Herman-JCP55-HW} which, for linear molecules, 
is defined as
\begin{eqnarray} \label{eq:EW}
 &F_{RP} = [1 + A_1 m + A_2^{RP}\,m^2]^2 \quad 
    &\text{for $P$ and $R$ branch lines} \,, \notag \\
 &F_Q = [1 + A_2^Q J(J + 1)]^2           \quad 
    &\text{for $Q$ branch lines} \notag  \,, \notag
\end{eqnarray}
where $m = -J$ and $m = J + 1$ for the $P$ and $R$ branches, respectively.
The $A_n^{PQR}$ coefficients depend on the quadratic and cubic potential constants and 
express the small effects due to the molecular non-rigidity on the intensity factors
\citep{Watson-JMS87-HW}.
For regular, semi-rigid molecules, the contribution of the Herman--Wallis factor to 
$\mathfrak{R}$ is of only a few percent for high $J$ values \citep[see, e.g.][]{ElHach-JMS02-C2H2}.
However, they should be considered when high-accuracy in the relative intensity calculation
is required.

\subsection{Astrophysical Implications} \label{sec:disc:astro}
\indent\indent
The improved set of molecular data presented here provides a useful guidance for the searches 
of \hctn in extra-terrestrial environments and may help to retrieve accurate quantitative 
information from the observations.
We obtained an improved description of the ro-vibrational energies that includes a careful
modelling of various local spectral perturbations.
This achievement is beneficial to the IR studies of planetary atmospheres (e.g., Titan),
whose outcome relies on the ability of predicting accurately the hot bands intensity 
distribution \citep[e.g.,][]{Jolly-JMS07-HC3N,Jolly-APJ10-HC4H}.
More importantly, it helps in interpreting the crowded millimetre spectra observed towards 
some chemically rich regions of the ISM.

\citet{Belloche-AA16-EMoCA} have recently published a complete 3\,mm spectral survey of 
the hot molecular core Sgr~B2(N2) performed with the Atacama Large Millimeter/submillimeter 
Array (ALMA). 
This survey, called ``Exploring Molecular Complexity with ALMA'' (EMoCA), resulted in the 
detection of a number of complex organic molecules, including \hctn in the ground as well 
as in many vibrationally excited states (see their Table~4). 
The LTE modelling of the full \hctn spectral profile proved to be successful, with some 
inconsistencies at 92.1\,GHz and 100.4\,GHz due to the incorrect predictions for the pair 
of interacting states $\varv_5 = \varv_7 = 1$ and $\varv_4 = 7$.
In fact, these frequencies correspond to $J = 9,10$ lines where the crossing between 
$0^f$ and $2^f$ sublevels of the above mentioned states occurs, hence the corresponding 
transitions are considerably displaced from their hypothetically unperturbed positions.

%%%%% FIGURE EMOCA ...
\begin{figure*}[t!]
  \centering
  \includegraphics[angle=0,width=0.8\textwidth]{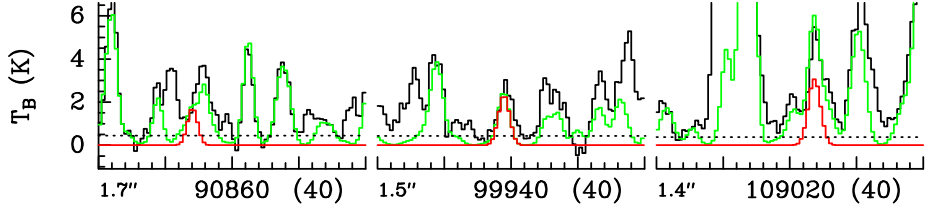}
  \caption{Transitions of HC$_3$N, $\varv_4 = 1$ covered by the EMoCA survey. 
           The best-fit LTE synthetic spectrum of \hctn, $\varv_4 = 1$ is displayed in 
           red and overlaid on the observed spectrum of Sgr~B2(N2) shown in black. 
           The green synthetic spectrum contains the contributions of all molecules 
           identified in the survey so far, including the species shown in red. 
           The central frequency and width are indicated in MHz below each panel. 
           The angular resolution (HPBW) is also indicated. 
           The $y$-axis is labeled in brightness temperature units (K). 
           The dotted line indicates the $3\sigma$ noise level.}
  \label{fig:emoca_v4e1}
\end{figure*}

%%%%% FIGURE EMOCA ...
\begin{figure*}
  \centering
  \includegraphics[angle=0,width=0.8\textwidth]{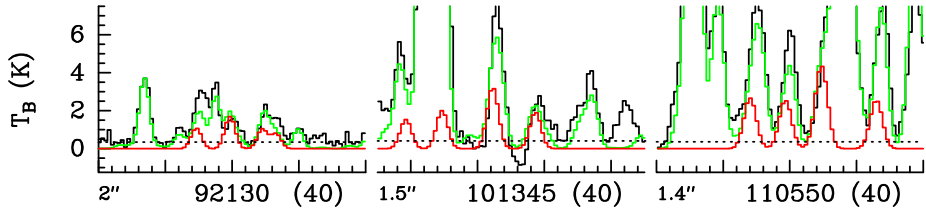}
  \caption{Same as Figure~\ref{fig:emoca_v4e1} for $\varv_7 = 4$.}
  \label{fig:emoca_v7e4}
\end{figure*}

%%%%% FIGURE EMOCA ...
\begin{figure*}
  \centering
  \includegraphics[angle=0,width=0.8\textwidth]{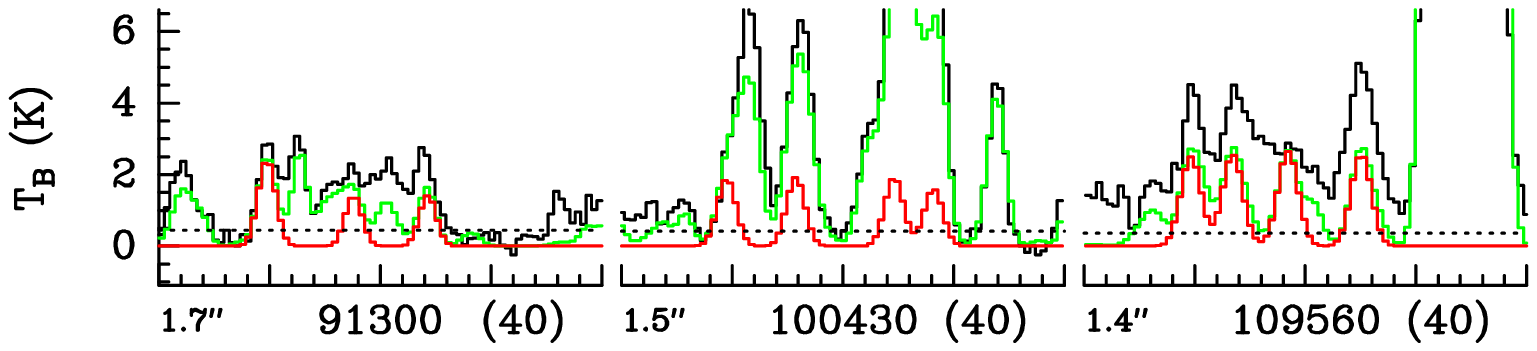}
  \caption{Same as Figure~\ref{fig:emoca_v4e1} for $\varv_5 = \varv_7 = 1$.} 
  \label{fig:emoca_v5e1v7e1}
\end{figure*}

\begin{figure}[t!]
  \centering
  \includegraphics[angle=0,width=0.8\textwidth]{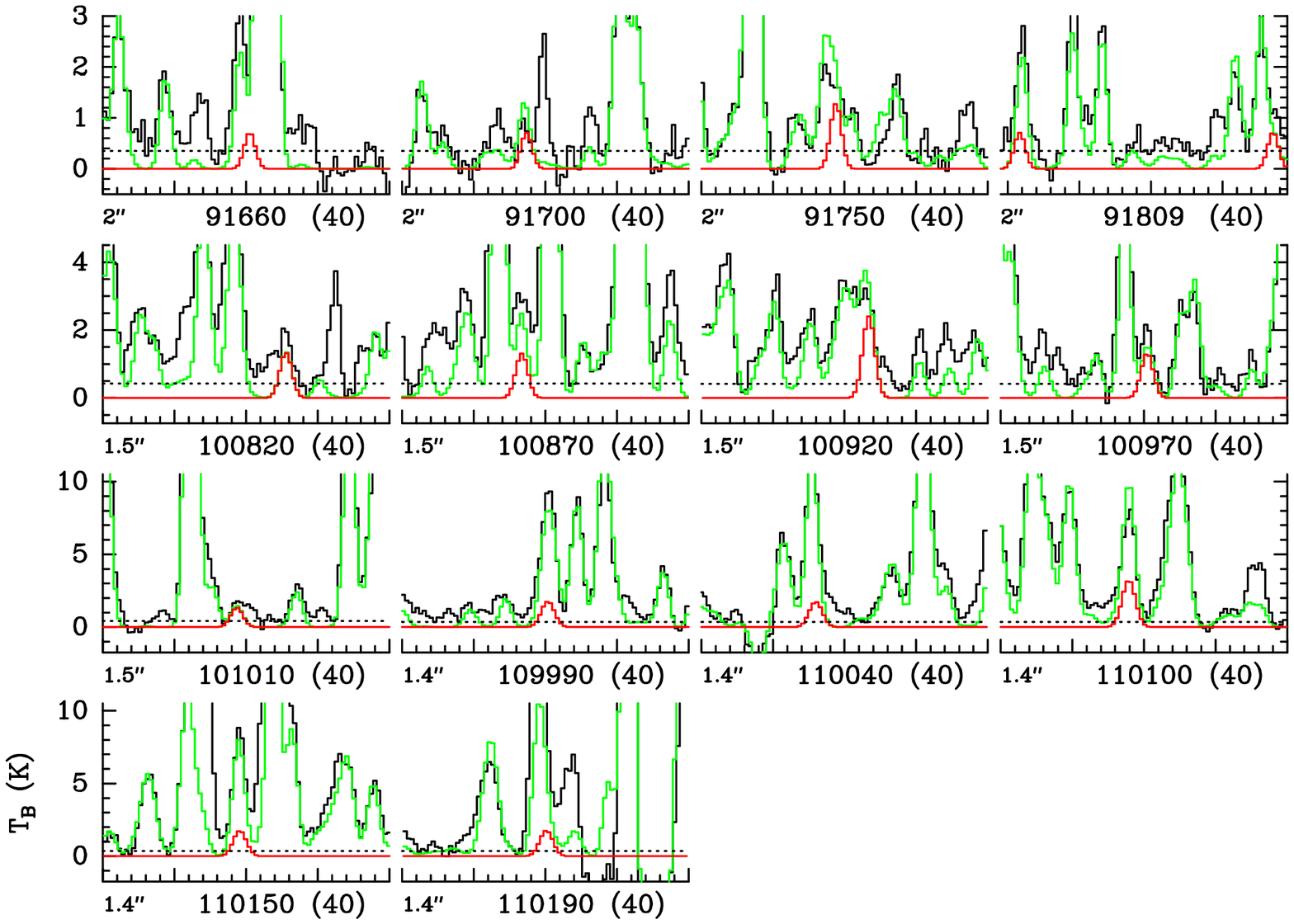}
  \caption{Same as Figure~\ref{fig:emoca_v4e1} for $\varv_6 = 1$, $\varv_7 = 2$.}
  \label{fig:emoca_v6e1v7e2}
\end{figure}

\begin{figure}
  \centering
  \includegraphics[angle=0,width=0.8\textwidth]{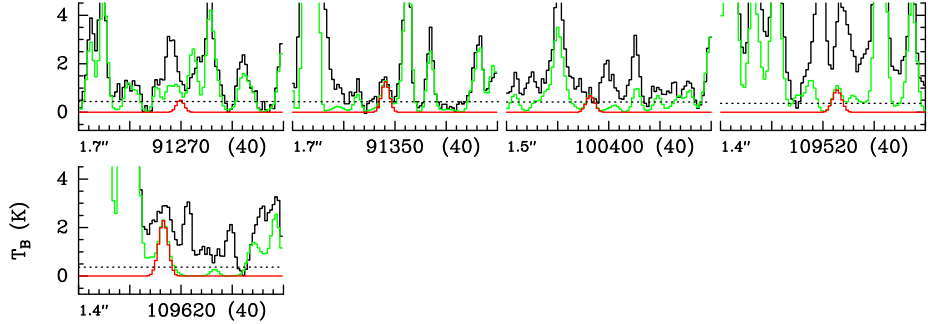}
  \caption{Same as Figure~\ref{fig:emoca_v4e1} for $\varv_6 = 2$.}
  \label{fig:emoca_v6e2}
\end{figure}

Here, we use our new spectroscopic predictions to revisit the analysis of the \hctn 
emission in the EMoCA spectrum of Sgr~B2(N2). 
We model the emission of the vibrationally excited states of \hctn above $\varv_7 = 1$ 
assuming local thermodynamic equilibrium (LTE), with the same parameters as derived in 
\citet{Belloche-AA16-EMoCA}: a source size of 0.9$''$, a rotational temperature of 200\,K, 
a linewidth of 5.8\,km\,s$^{-1}$, a velocity offset of $-1.0$\,km\,s$^{-1}$ with respect 
to the assumed systemic velocity of Sgr~B2(N2) of 74\,km\,s$^{-1}$, and a column density 
of $5.2 \times 10^{17}$\,cm$^{-2}$ \citep[see Sect.~5.3 of][]{Belloche-AA16-EMoCA}.
The computation has been performed using the software \textsc{Weeds} \citep{Maret-AA11-WEEDS} 
taking into consideration the spectral-window- and measurement-set-dependent angular 
resolution of the observations.
The Einstein's $A$ constant for each transition has been computed using the experimental 
values of the dipole moment derived by \citet{DeLeon-JCP85-HC3N}.
The resulting synthetic spectra for the states above 800\,cm$^{-1}$, 
$\varv_4 = 1$, $\varv_7 = 4$, $\varv_5 = \varv_7 = 1$, $(\varv_6 = 1,\varv_7 = 2$), and 
$\varv_6 = 2$, are overlaid on the ALMA spectrum of Sgr~B2(N2) in 
Figs.~\ref{fig:emoca_v4e1}--\ref{fig:emoca_v6e2}.

\begin{figure}
  \centering
  \includegraphics[angle=0,width=0.8\textwidth]{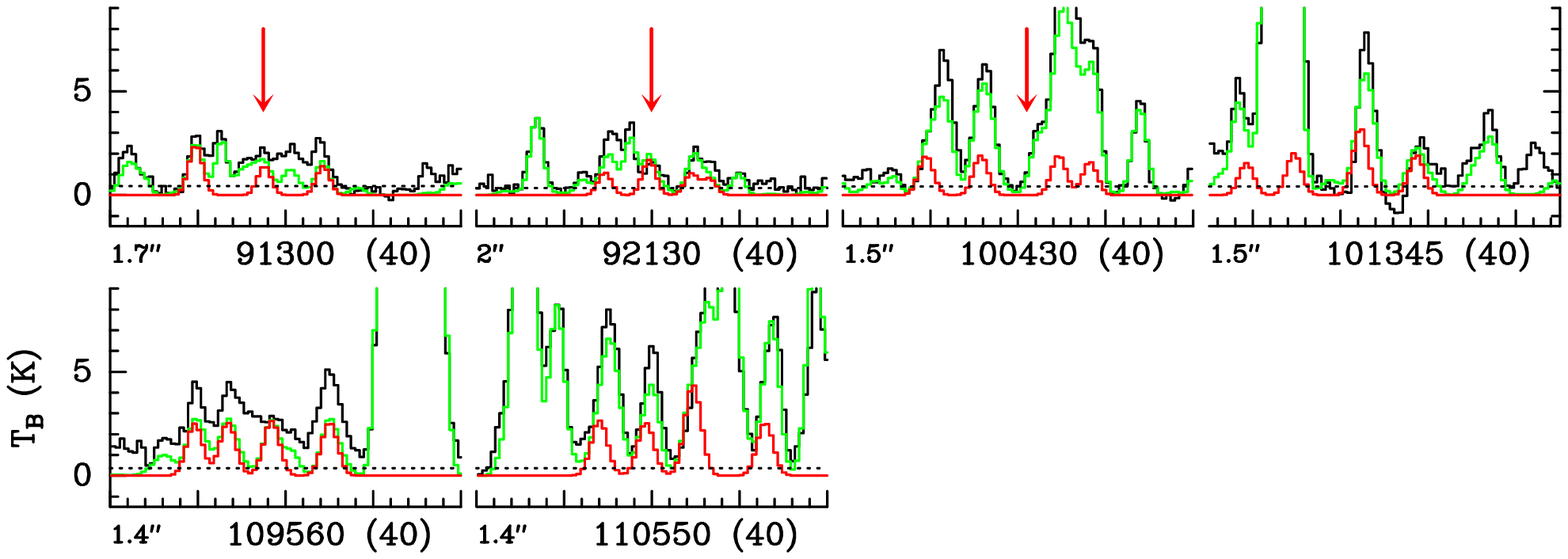}
  \caption{Same as Figure~\ref{fig:emoca_v4e1} for $\varv_7 = 4$ and $\varv_5=\varv_7=1$ 
           together in order to compare to older predictions shown in 
           Figure~\ref{fig:emoca_old}. 
           The arrows mark the frequencies where the new spectroscopic predictions improve 
           the agreement between the synthetic and observed spectra.}
  \label{fig:emoca_new}
\end{figure}

\begin{figure}
  \centering
  \includegraphics[angle=0,width=0.8\textwidth]{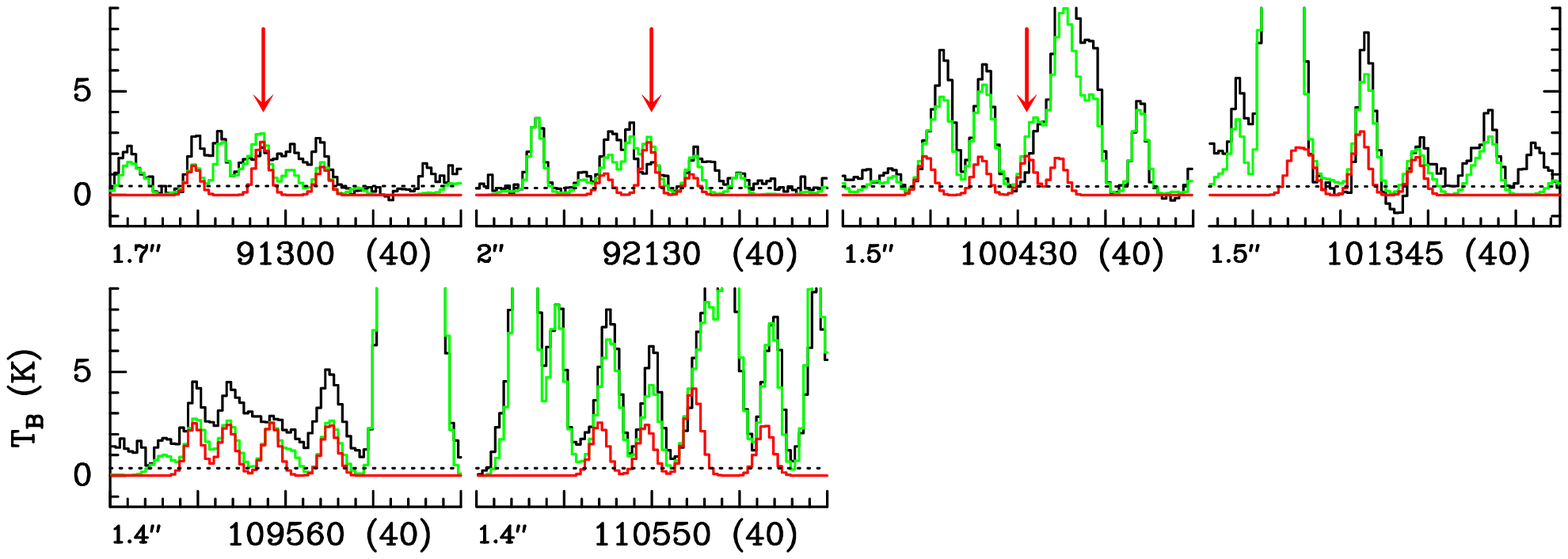}
  \caption{Same as Figure~\ref{fig:emoca_new} but using the older spectroscopic predictions 
           as in \citet{Belloche-AA16-EMoCA}. 
           The arrows mark the frequencies where discrepancies between the synthetic and 
           the observed spectra are present.}
  \label{fig:emoca_old}
\end{figure}

The model that uses our new spectroscopic predictions gives the same results as the older 
one for $\varv_4 = 1$ and $\varv_6 = 2$, but it improves the agreement between the 
synthetic and observed spectra for $\varv_7 = 4$ and $\varv_5=\varv_7=1$ thanks to the 
proper treatment of the interaction between states in the spectroscopic analysis. 
This is illustrated in Figs.~\ref{fig:emoca_new} and \ref{fig:emoca_old} which display 
synthetic models computed with the new and old spectroscopic predictions, respectively,
over the frequency ranges where $\varv_7 = 4$ and $\varv_5 = \varv_7 = 1$ have rotational 
transitions. 
In these figures, the arrows indicate the frequencies where the new predictions 
(Figure~\ref{fig:emoca_new}) solve inconsistencies that were present when using the older 
ones (Figure~\ref{fig:emoca_old}).

%%%%% FIGURE EMOCA ...

%%%%% FIGURE EMOCA ...

We also show in Figure~\ref{fig:emoca_v6e1v7e2} the synthetic rotational spectrum of the 
excited state $\varv_6 = 1$,$\varv_7 = 2$. 
Most of its transitions are unfortunately blended with transitions of other species in 
the ALMA spectrum of Sgr~B2(N2). 
There are, however, two transitions that suffer less from contamination and can be 
considered as detected (at 100826\,MHz and 100970\,MHz). 
Small discrepancies can be seen around 91697\,MHz and 100923\,MHz, where the model 
containing all the identified molecules slightly overestimates the observed spectrum, 
but these discrepancies are at the $2\sigma$ level only and we consider them as 
insignificant. 
A discrepancy at the $3\sigma$ level is present around 91748\,MHz. 
The identified emission is dominated by acetone in its $\varv_{12} = 1$ state which is 
also responsible for the discrepancy present at 91755\,MHz. 
Therefore we believe the discrepancy at 91748\,MHz is due to an inaccurate modelling
of the acetone spectrum and not to \hctn $\varv_6 = 1$,$\varv_7 = 2$. 
Finally, a discrepancy at the $10\sigma$ level is present around 110098\,MHz. 
Here again, the emission is dominated by acetone in its $\varv_{12}=1$ state, so we 
suspect that the discrepancy is due to an issue with our LTE model of acetone, or with
the spectroscopic predictions of this species. 
Therefore, all in all, we are confident that \hctn $\varv_6 = 1$,$\varv_7 = 2$ features 
are present at the level indicated by our synthetic spectrum. 
The detection of this excited state was not reported in \citet{Belloche-AA16-EMoCA} due 
to the lack of spectroscopic predictions at the time their study was published.

\section{Conclusion} \label{sec:conc}
\indent\indent
Cyanoacetylene is a molecule of remarkable astronomical importance and has been observed 
in a number of sources, both Galactic and extra-galactic. 
These detections relied on laboratory investigations which, however extended, lacked some 
essential information concerning the rotational and ro-vibrational spectra. 
Indeed, the knowledge of the fundamentals and of the weak hot bands involved in the IR 
spectrum, necessary to model the molecular profile of planetary atmospheres, was incomplete. 
Moreover, the pure rotational spectrum of \hctn observed in Space sometimes could not 
be assigned successfully because of the density of lines or the incorrect predictions 
based on laboratory analyses. 

This work aims at filling these gaps by undertaking a full re-investigation of the IR 
spectrum of \hctn up to 1100\,\wn  by high-resolution FTIR spectroscopy. 
In addition, several pure rotational transitions in the ground and vibrationally-excited 
states have been recorded in the mm and submm regions. 
In total, all the transitions present in the literature and newly-recorded in this work, 
involving energy levels below 1000\,\wn, form a data set of about~3400 ro-vibrational 
lines across 13~bands and~1500 pure rotational lines belonging to 12~vibrational states. 
They have been fitted together to an effective Hamiltonian allowing the determination 
of 121~spectroscopic constants.   
Such a global data analysis could not be accomplished without considering explicitly the 
complex network of vibrationally-interacting states. 

In the energy interval considered at present are two major resonance schemes: 
$i$) $\varv_5 = 1$ $\sim$ $\varv_7 = 3$, 
$ii$) $\varv_4 = 1$ $\sim$ $\varv_5 = \varv_7 = 1$ $\sim$ $\varv_6 = 2$ 
$\sim$ $\varv_7 = 4$\@. 
The interaction terms of the Hamiltonian are purely vibrational ($\oH_{30}$, $\oH_{40}$, 
$\oH_{50}$) and ro-vibrational ($\oH_{32}$, $\oH_{42}$, $\oH_{52}$)\@.
The isolated states are the ground state, $\varv_6 = 1$, $\varv_7 = 1$, $\varv_7 = 2$, 
$\varv_6 = \varv_7 = 1$, and $\varv_6 = 1,\varv_7 = 2$. 
The energy cutoff of 1000\,\wn was chosen so that a complete analysis of the low-lying 
vibrational states involved in the anharmonic resonances could be performed. 
Transitions involving higher energy levels, although detected in our experiments, have 
not been considered in the present study.
Some of these higher-level transitions are part of the same interactions but are simply 
scaled up by one vibrational quanta of $\varv_7$.
Rotational, vibrational and resonance constants have been determined from the global 
fit without any assumption deduced from theoretical calculations or through comparisons to 
similar molecules. 
The overall quality of the fit is very satisfactory and the parameters have been derived 
with very good precision and accuracy. 
Eventually, it was possible to compute a large set of reliable accurate ro-vibrational 
rest frequencies for all the vibrational levels of \hctn below 1000\,\wn and for pure 
rotational transitions in the $J$-range between~0 and~120\@. 
This is particularly important for the spectral regions not explored in laboratory. 
Our predictions, which form the most accurate and complete set of rest frequencies 
available for \hctn, are especially useful for astronomical searches. 

These improved spectral predictions have enabled refined analyses of molecular
emission observed towards Sgr~B2(N2) with ALMA (EMoCA survey).
Discrepancies between observations and the global model \citep{Belloche-AA16-EMoCA}, 
produced by perturbed \hctn ro-vibrational lines, could be effectively removed.
Furthermore, one previously unreported vibrational state of \hctn 
($\varv_6=1,\varv_7=2$) has been newly identified in the EMoCA observed spectra.

\section*{Acknowledgments}
\indent\indent
The authors acknowledge the financial support of the Ministero dell'Istruzione, 
dell'Università e della Ricerca (PRIN 2012 funds, project STAR) [grant number 20129ZFHFE] 
and the University of Bologna (RFO funds).
H.S. thanks Andr\'e Fayt for initial prediction which were beneficial for the study of 
higher lying vibrational states of \hctn.
This paper makes use of the following ALMA data: ADS/JAO.ALMA\#2011.0.00017.S, 
ADS/JAO.ALMA\#2012.1.00012.S. 
ALMA is a partnership of ESO (representing its member states), NSF (USA), and 
NINS (Japan), together with NRC (Canada), NSC and ASIAA (Taiwan), and KASI 
(Republic of Korea), in cooperation with the Republic of Chile. 
The Joint ALMA Observatory is operated by ESO, AUI/NRAO, and NAOJ. 
The interferometric data are available in the ALMA archive at \\ 
\texttt{https://almascience.eso.org/aq/}.
The work in Cologne and Bonn has been in part supported by the Deutsche 
Forschungsgemeinschaft (DFG) through the collaborative research grant SFB~956 
``Conditions and Impact of Star Formation'', project area B3.
Part of the early laboratory studies in Cologne were supported through SFB~494.

%%%%%%%%%%%%%%%%%%%%%%% REFERENCES
%% Dal listone personale
%\bibliographystyle{aa}
%\bibliography{/home/luca/Lavoro/bibbase/jsc-astro,%
%              /home/luca/Lavoro/bibbase/cchains,%
%              /home/luca/Lavoro/bibbase/diatomic,%
%              /home/luca/Lavoro/bibbase/poly-ions,%
%              /home/luca/Lavoro/bibbase/astroch,%
%              /home/luca/Lavoro/bibbase/molphys,%
%              /home/luca/Lavoro/bibbase/instrument,%
%              /home/luca/Lavoro/bibbase/misc}

\end{document}